\documentclass[aps,prl,twocolumn,showpacs,10pt,superscriptaddress,floatfix,longbibliography]{revtex4-2}

\usepackage{newtxtext,newtxmath}
\usepackage{graphicx}
\usepackage{subfigure}
\usepackage{amsmath}
\usepackage{amsfonts}
\usepackage{mathrsfs}
\usepackage{bbm}
\usepackage{subfigure}
\usepackage{xcolor}
\usepackage[colorlinks=true]{hyperref}
\usepackage{ulem}
\usepackage[T1]{fontenc}

\newcommand{\blue}[1]{\textcolor{blue}{#1}}

\begin{document}

\title{Transport theory for topological Josephson junctions with a Majorana qubit}

\author{Zhi Wang}
\altaffiliation{These authors contributed equally to this work.}
\affiliation{School of Physics, Sun Yat-sen University, Guangzhou 510275, China}

\author{Jia-Jin Feng}
\altaffiliation{These authors contributed equally to this work.}
\affiliation{International Center for Quantum Materials, Peking University, Beijing 100871, China}

\author{Zhao Huang}
\email{Corresponding author: huangzhaophysics@gmail.com}
\affiliation{Texas Center for Superconductivity, University of Houston, Houston, Texas 77204, USA}

\author{Qian Niu}
\affiliation{School of Physics, University of Science and Technology of China, Hefei, Anhui 230026, China }

\begin{abstract}
We construct a semiclassical theory for the transport of topological junctions starting from a microscopic Hamiltonian that comprehensively includes the interplay among the Majorana qubit, the Josephson phase, and the dissipation process. With the path integral approach, we derive a set of semiclassical equations of motion that can be used to calculate the time evolution of the Josephson phase and the Majorana qubit. In the equations we reveal rich dynamical phenomena such as the qubit induced charge pumping, the effective spin-orbit torque, and the Gilbert damping. We demonstrate the influence of these dynamical phenomena on the transport signatures of the junction.
We apply the theory to study the Shapiro steps of the junction, and find the suppression of the first Shapiro step due to the dynamical feedback of the Majorana qubit.
\end{abstract}
\maketitle

{\it \blue { Introduction.}}-- Josephson physics receives reviving interests due to the rapid progress of superconducting quantum computation in recent years\cite{devoret2013review,martinis2020}. The demand for the minification of the superconducting quantum circuits pushes the limits of the size of Josephson junctions\cite{kjaergaard2019,siddiqi2021}. For junctions that are small enough, a single embedded qubit may significantly modify the transport signatures\cite{tinkham2004}. While this effect has been discussed in a number of systems with various models\cite{feng2018hysteresis,choi2020,oriekhov2021}, a comprehensive theory that takes account the qubit dynamics is still absent. 

This issue is particularly relevant to topological Josephson junctions with Majorana zero modes\cite{deng2012anomalous,oostinga2013,peng2016TJJ,cayao2017,deacon2017radiation,cayao2017,kamata2018,schrade2018,lei2018prl,li2018prb,liu2019PRL,laroche2019radiation,ren2019josephson,fornieri2019,he2019,stern2019,klees2020,avila2020,razmade2020,scharf2021,dartiailh2021,li2021prl,jian2021,zhang2021prl}. The two Majorana zero modes in the junction construct a Majorana qubit which results in a $4\pi$-periodic Josephson current\cite{kitaev2001unpaired,Kwon2004}. Previous theoretical studies take a variety of phenomenological models which are different extensions of the standard resistively shunted junction model for conventional junctions
\cite{dominguez2012,dominguez2017,feng2018hysteresis,choi2020,svetogorov2020,frombach2020}. However, these phenomenological models still have difficulties in explaining the experimental reported transport features such as the suppression of the first Shapiro step\cite{rokhinson2012fractional,wiedenmann2016shapiro4pi,bocquillon2017gapless,li2018NM,wang2018Shapiro,Schuffelgen2019,LeCalvez2019,rosenbach2021}. It is highly desirable to construct a microscopic theory to examine the validity of the phenomenological models and to understand the experimental results.

In this work, we develop a semiclassical theory for studying the transport properties of the topological junctions. Our theory starts from a microscopic Hamiltonian that characterizes the coupling between the Josephson junction and the Majorana qubit. We take a path integral approach to incorporate the dissipation process that is essential for studying the transport properties, and derive the semiclassical equations of motion for the Josephson phase and the Majorana qubit. In the equations of motion, we identify the effective spin-orbit torque and the Gilbert damping in the qubit dynamics, and reveal the charge pumping driven by the qubit rotation in the dynamics of the Josephson phase. Solving the equations of motion, we obtain the time evolution of the Josephson phase which provides transport and spectroscopic signatures for the junction.

As an application of this theory, we calculate the Shapiro steps of the topological junction. We find that the first step is strongly suppressed while higher odd-number steps are robustly visible for a range of junction parameters. 
We show that this bizarre behavior is due to the feedback of the Majorana qubit dynamics to the transport of the junction. At the voltage of the first step, the Majorana qubit evolves to a stable state which supports a finite $4\pi$-period Josephson current, and this $4\pi$-periodicity in the Josephson phase dynamics suppresses the first Shapiro step. At voltages of higher odd-number steps, however, the Majorana qubit evolves to a stable state which contributes a vanishing $4\pi$-period Josephson current, and the Shapiro steps are naturally intact.
Our theory provides an intrinsic mechanism for the reported Shapiro step missing in topological junctions.

\begin{figure}[b]
\begin{center}
\includegraphics[clip = true, width =\columnwidth]{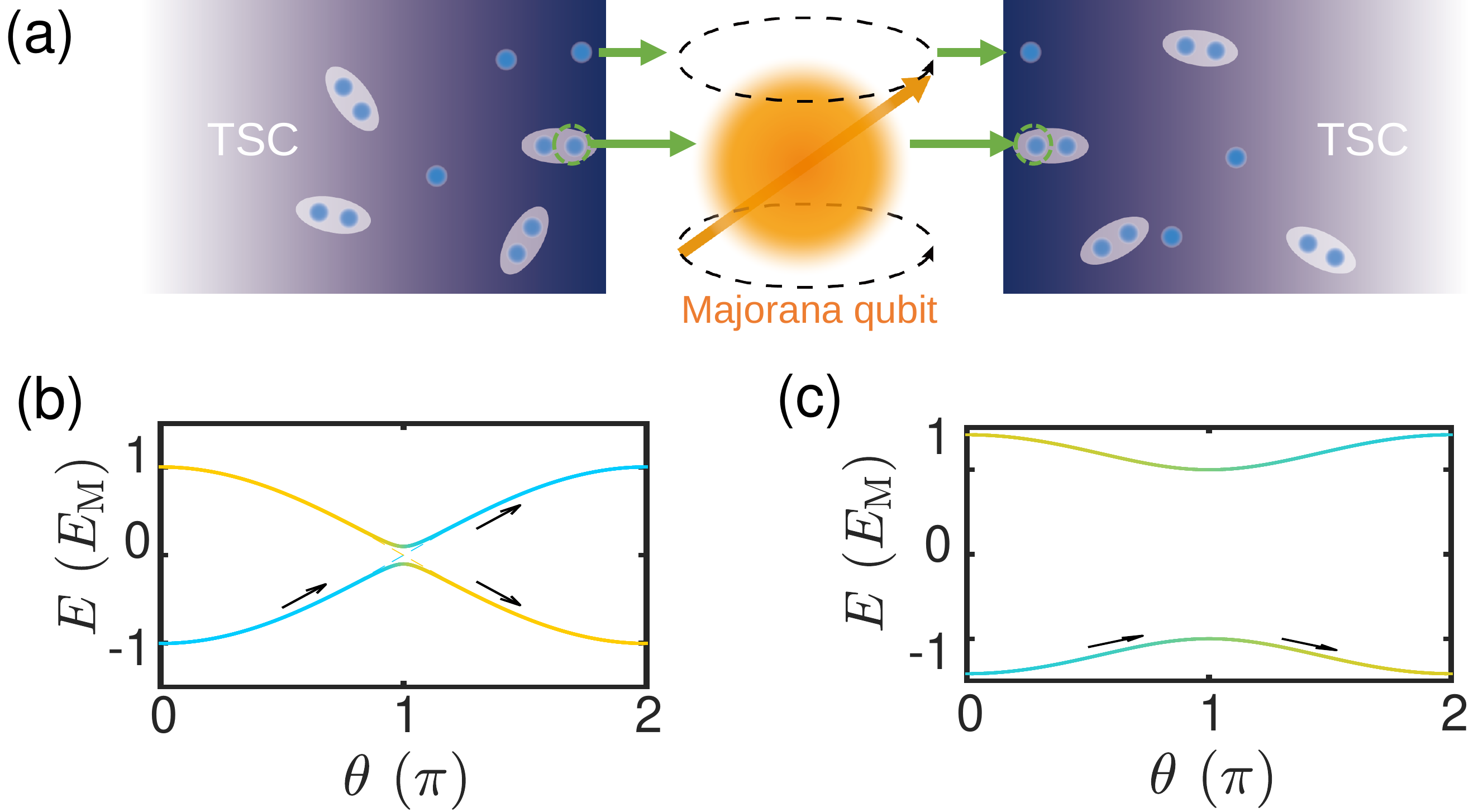}
\end{center}
\caption{(a) Schematic illustration of the two tunneling processes through the Majorana qubit: The qubit assisted half-pair tunneling that leads to the fractional Josephson effect, and the qubit rotation induced charge pumping. (b) The energy levels of the Hamiltonian ${\bf h} \cdot {\hat \sigma}$. Landau-Zener transitions happen at the anti-crossing points at $\theta = (2n + 1)\pi$.  (c) The energy levels of the Hamiltonian $({\bf h}+\dot \theta {\bf B}_f )  \cdot {\hat \sigma}$ for high voltage of $\dot \theta { B}_f \gg E_{\rm M}$, where ${\bf B}_f = { B}_f \hat x$. The Landau-Zener transition at the anti-crossing points are significantly suppressed and the qubit dynamics would follow one of the levels. }
\label{fig:model}
\end{figure}

{\it {\color{blue} Microscopic Hamiltonian and equations of motion.}}--The low-energy effective Hamiltonian for the junction with a Majorana qubit can be written as\cite{supp},
\begin{equation}\label{eq:JosephsonH}
\mathcal{H}_{\rm J}=\frac{\hat{p}^{2}_{\theta}}{C_{0}}-E_{\rm J}\cos \hat \theta - I_{\rm ex} \hat \theta- E_{\rm M}  \sigma_z \cos \frac {\hat \theta}{2}   + E'_{\rm M}  \sigma_x,
\end{equation}
where $\hat \theta$ is the Josephson phase with canonical momentum $\hat { p}_{\theta}=2e\partial_{\theta}$, $E_{\rm J}$ is the Josephson energy, $C_0=2C \hbar^{2}/\left(2e\right)^{2}$ is a dimensionless constant determined by the effective capacitance $C$ of the junction, $I_{\rm ex}$ represents the experimentally controllable external current injected into the junction, ${\sigma}_{x,z}$ are Pauli matrices which represent the pseudo-spin direction for the Majorana qubit, $E_{\rm M}$ and $E'_{\rm M}$ represent the energies of the Majorana qubit from various couplings between Majorana zero modes.
The first three terms of the Hamiltonian have been widely adopted for studying conventional Josephson junctions\cite{tinkham2004}, while the last two terms come from the Majorana zero modes\cite{kitaev2001unpaired,kitaev2003anyon} and can be derived from the Bogoliubov- de Gennes Hamiltonian of a topological Josephson junction\cite{lutchyn2010TSC1d,oreg2010TSC1d}.

This Hamiltonian can be understood as describing a spin-one-half {\it particle} with a mass of $C_0$, moving under a potential energy $U_p = -E_{\rm J}\cos\theta- I_{\rm ex} \theta$, and a Zeeman energy $U_z ={\bf h}\cdot{ \hat \sigma}$ where the direction of the Zeeman field ${\bf h}=\left(E'_{\rm M},0,-E_{\rm M}\cos\frac{\theta}{2}\right)$ varies along the path of the motion. The potential energy is identical to the tilted-washboard potential that was taken in studying conventional junctions\cite{tinkham2004}, while the unique Zeeman energy comes from the coupling between the Josephson phase and the Majorana qubit.

The time-evolution of the Josephson phase determines the transport properties of the junction through the ac Josephson relation\cite{josephson1962}. To derive the equation of motion for this time-evolution, we rewrite the Hamiltonian of Eq. (\ref{eq:JosephsonH}) into an action\cite{wen2004,Altland2010condensed,supp},
\begin{equation}\label{eq:JosephsonA}
S_{\rm J}  =\int dt \left(\frac{C_0}{2}\dot{\theta}^{2}+E_ {\rm J}\cos\theta+ I_{\rm ex} \theta+{\bf A}_{s}\cdot{ \dot{\bf s}}+{\bf h}\cdot{\bf s}\right),
\end{equation}
where  ${{\bf s}} =\psi^{\dagger}\mathbf{\hat{{\bf \sigma}}}\psi= (\sin \varphi \sin \phi, \sin \varphi \cos \phi,\cos \varphi)$ represents the psuedo-spin state on the Bloch sphere with $\psi=(e^{-i\phi}\cos\frac{\varphi}{2},\sin\frac{\varphi}{2})$ the spinor wave function of the qubit, ${\bf {\bf A}}_{s}={\hat {\bf e}_{\phi}} (1-\cos\varphi) / \sin\varphi$
represents the Berry connection on the Bloch sphere\cite{Altland2010condensed}, which provides a
Berry curvature of $\nabla\times{\bf A}_{s}={\bf s}$. The extreme action path of Eq. (\ref{eq:JosephsonA}) gives the semiclassical equations of motion for the Josephson phase $C_0 \ddot{\theta} + I_{c1}\sin\theta+I_{c2}s_{z}\sin\frac{\theta}{2} - I_{\rm ex} = 0$, and the pseudo spin $\dot{{\bf s}} ={\bf h} \times{\bf s}$,
where $I_{c1} =2e E_{\rm J} / \hbar $ is the supercurrent from the Cooper pair tunneling and $I_{c2} = e E_{\rm M}/\hbar$ is the supercurrent from the half-pair tunneling through the Majorana qubit.
The equations of motion explicitly demonstrate the coupling between the Josephson phase and the Majorana qubit through the $s_z$ dependent Zeeman term in the first equation and the $\theta$ dependent effective magnetic field in the second equation. However, these equations are inadequate for studying the transport properties of the junction. The missing piece is the dissipation process.

To include the dissipation into the equations of motion, we follow the Caldeira-Leggett approach and introduce a thermal bath of harmonic modes to characterize the environment\cite{Caldeira1981TLS,Caldeira1983Brownian}. The environmental degrees of freedom and their coupling with the junction can be described with the action\cite{supp},
\begin{equation}\label{eq:environmentA}
S_{\rm en}  =\sum_{i} \int dt \left(\frac{1}{2}\left(\dot{h}_{i}^{2}-\Omega_{i}^{2}h_{i}^{2}\right) + h_{i}\left(g_{i}\theta+{\bf B}_{i}\cdot{\bf s}\right)\right),
\end{equation}
where $h_{i} $ are the coordinates of the environmental modes and $\Omega_{i}$ are
their oscillating energies, 
$g_i$ represents the minimal coupling between the environmental modes and the Josephson phase\cite{Caldeira1983Brownian}, and ${\bf B}_i$ represents the minimal coupling between the environmental modes and the qubit\cite{Leggett1987RMP}. The details of $g_i$ and ${\bf B}_{i}$ are determined by the coupling between each environmental mode and the junction. For topological junctions described by Eq. (\ref{eq:JosephsonH}), the environmental modes that modulate the tunneling barrier of the junction can be understood
as a fluctuation on the amplitude of $E_{\rm M}$ and we have ${\bf B}_{i}=B_{i}\hat{z}$, while the
environmental modes that modulate the coupling between Majorana zero modes in one side of the junction can be understood as a fluctuation on $E'_{\rm M}$ and we have ${\bf B}_{i}=B_{i}\hat{x}$. In general, the environment modes may originate from multiple sources, such as electromagnetic perturbations, lattice vibrations, as well as thermally activated electrons. These modes could have much more complex couplings with the junction, while the Caldeira-Leggett model is the minimal model for describing the dissipation processes.

The dissipated evolution of the junction can be obtained by integrating out the environmental degrees of freedom. This is achievable since the environment is modeled with harmonic modes. After the integration, we arrive at an effective action for the junction variables\cite{supp},
\begin{align}
S_{\rm eff} &=\int dt \left(\frac{C_0 }{2}\dot{\theta}^{2}+E_ {\rm J}\cos\theta+ I_{\rm ex} \theta+{\bf A}_{{ s}}\cdot { \dot{\bf s}}+{\bf h} \cdot{\bf s}\right)
\\ \nonumber
 &+\frac{1}{4}\int dtdt'\left[\eta_{\alpha}(t)-\eta_{\alpha}(t')\right]G_{\alpha\beta}(t,t')\left[\eta_{\beta}(t)-\eta_{\beta}(t')\right],
\end{align}
where $\eta_\alpha=(\theta,s_{x},s_{y},s_{z})$ represents the junction degree of freedom, ${G}_{\alpha \beta}(t,t')  =-i\frac{\tilde{M}_{}}{\pi}\frac{1}{|t-t'|^{2}}$ is the averaged Green function from the environmental modes, with $\tilde{M}$ the averaged coupling matrix\cite{supp}.
The least action path for this effective action provides the full semiclassical equations of motion for the junction variables,
\begin{subequations}\label{eq:EOM}
\begin{align}
I_{\rm ex} & = C_0 \ddot{\theta}+\frac{\dot{\theta}}{R}+I_{c1}\sin\theta+ I_{c2}s_{z}\sin\frac{\theta}{2}+{\bf B}_{f}\cdot\dot{{\bf s}}, \label{eq:EOM1}\\
\dot{{\bf s}}  & ={\bf h\times{\bf s}}+\dot{\theta}{\bf B}_{f}\times{\bf s}+\left(\tilde{\gamma}\cdot\dot{{\bf s}}\right)\times{\bf s},\label{eq:EOM2}
\end{align}
\end{subequations}
where $R = 1/\text{\ensuremath{\sum_{i}}}g_{i}^{2}$ is the effective resistance of the junction which comes from the coupling between the environment and the Josephson phase, ${\bf B}_{f} =\text{\ensuremath{\sum_{i}}} g_i {\bf B}_i$ is the environment mediated coupling field between the Josephson phase and the qubit, and $\tilde{\gamma}_{\alpha \beta} =\sum_{i} { B}_{i\alpha} B_{i\beta}$ represents the environment induced dissipation to the qubit. 

The equation (\ref{eq:EOM}) is the central result of this work. Solving these two self-consistent equations we can obtain $\theta (t)$ which determines the dc and ac voltage of the junction through the Josephson relation $V(t) = \hbar {\dot \theta}(t)/2e$. Starting from a microscopic Hamiltonian, we provide a framework to study the transport properties of a Josephson junction with an embedded Majorana qubit.

{\it {\color{blue} Physical interpretation of the equations of motion.}}-- Let us setup a physical picture for interpreting the terms in the equations of motion, particularly those terms coming from the embedded Majorana qubit. The equation (\ref{eq:EOM1}) is a current conservation equation which states that the externally injected current $I_{\rm ex}$ equals the current flowing through all the physical channels in the junction. If the qubit is completely ignored, then the last two terms on the right side of the Eq. (\ref{eq:EOM1}) should be dropped out and the equation becomes a self-consistent equation for the Josephson phase. This is exactly the resistively and capacitively shunted junction model that has been widely used for studying bulk Josephson junctions\cite{tinkham2004}.

The qubit provides two additional terms in Eq. (\ref{eq:EOM1}) as illustrated in Fig. \ref{fig:model}a. The first term is the $4\pi$-periodic Josephson current $I_{c2}s_{z}\sin\frac{\theta}{2}$, which is linearly dependent on the z-component of the pseudo spin. This is the extensively discussed fractional Josephson effect\cite{kitaev2001unpaired,lutchyn2010TSC1d,alicea2011non}, which comes from the qubit assisted half-pair tunneling in the junction. The other term is ${\bf B}_{f}\cdot\dot{{\bf s}}$ which is non-vanishing only when the pseudo spin rotates. Since the pseudo-spin state of the Majorana qubit is defined by the parity of the superconducting ground state, this current represents the pumped current by the parity flipping of the Majorana qubit. This qubit pumping has never been revealed in previous models, and only becomes apparent from the effective action of the microscopic theory.

Now we take a closer look at the Eq. (\ref{eq:EOM2}). In the absence of the environment, only the first term on the right side of equation survives. The residing equation, $\dot{{\bf s}} = {\bf h\times{\bf s}}$, describes a qubit procession where the direction of the procession $\bf h$ oscillates with $\theta$. When the oscillating component $h_z$ is much larger than the stable component $h_x$, the qubit would evolve under an oscillating energy spectrum shown in Fig. \ref{fig:model}b. When $\theta$ moves through the anti-crossing points $\theta = (2n + 1)\pi$, the qubit experiences Landau-Zener transitions, and multiple coherent Landau-Zener transitions exhibit St\"{u}ckelburg interference\cite{shevchenko2010LZS}. These effects have been thoroughly analyzed in previous phenomenological models\cite{feng2018hysteresis,feng2020radiation}.

The second term on the right side of Eq. ({\ref{eq:EOM2}}) is a unique discovery of our theory. It resembles a spin-orbit torque which linearly depends on the velocity of the Josephson phase.
This spin-orbit torque dominates the qubit dynamics at high voltage of $\dot \theta \gg E'_{\rm M}/\hbar$, causing a significant suppression of the energy crossing and the Landau-Zener transition, as shown in Fig. \ref{fig:model}c. For this reason, the transport and spectroscopic signals of the junction are expected to exhibit qualitatively different behaviors for different voltage regimes. This is useful for understanding the voltage-dependent signatures that have been widely reported in the I-V characteristics curves and Josephson radiations of Josephson junctions constructed by topological systems\cite{oostinga2013,deacon2017radiation,laroche2019radiation}.

The third term on the right side of Eq. ({\ref{eq:EOM2}}) is the anisotropic Gilbert damping which determines the dissipation of the qubit from the coupling to the environment. For isotropic case where the matrix $\tilde{\gamma}$ becomes a number, this term turns into the standard Gilbert damping which appeared in the Landau-Lifshiz-Gilbert equation\cite{Gilbert2004}. While the Gilbert damping has been widely taken to study the dynamics of the magnetization, our work provides a derivation for its microscopic origin in the Majorana qubit. This damping process influences the dynamics of the qubit and thereby modifies the transport properties of the junction.

Finally, we hope to point out that if the environment mediated coupling ${\bf B}_f$ and the Gilbert damping $\tilde \gamma$ are ignored, then Eq. (\ref{eq:EOM}) will reduce to the phenomenological quantum resistively and capacitively shunted junction model that has been taken to study the I-V characteristics and the Josephson radiations of the topological junction\cite{feng2018hysteresis,feng2020radiation}. Our microscopic theory clarifies the validity and limits of the phenomenological model.

\begin{figure}[tb]
\begin{center}
\includegraphics[clip = true, width =0.9 \columnwidth]{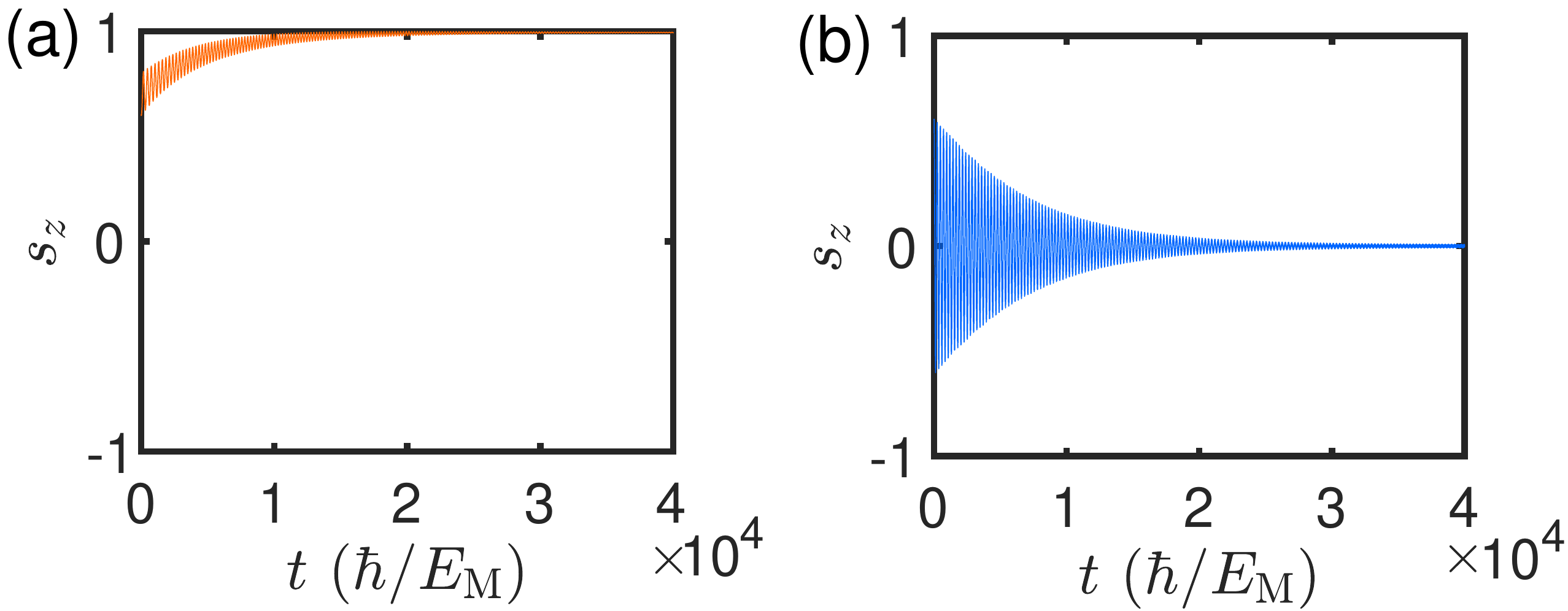}
\end{center}
\caption{ Typical time evolution of the Majorana qubit for the low voltage regime (a) and the high voltage regime (b). The damped oscillation are combined effect of the Landau-Zener-St\"{u}ckelburg interference and the Gilbert damping.
The two different stable values of the $s_z$ represent the different fixed points in the dynamics of the Majorana qubit.}
\label{fig:FX}
\end{figure}

{\it {\color{blue} Fixed point analysis.}}-- The Eq.~(\ref{eq:EOM}) are complicate nonlinear equations for which obtaining analytical solutions is impossible. However, the fixed points of the equations can be analytical calculated with the method of averaging, which is a method to decouple the nonlinear equations with the division of the dynamical variables to the "fast variables" and the "slow variables" based on their time scales\cite{sanders2007averaging}. In Eq.~(\ref{eq:EOM}) we treat the psuedo spin $\bf s$ as the slow variable since it has a larger time scale. We take it as constant to solve the Eq.~(\ref{eq:EOM1}) for the fast variable $\theta(t)$, and the solution provides the time-averaged Josephson energy $\int dt E_{\rm M} \cos \theta(t)/2  \approx \alpha s_z E_{\rm M}$. Plugging this into the Eq. (\ref{eq:EOM2}), we obtain an approximated self-consistent equation for $\bf s$, and the fixed points of this equation can be determined analytically. There are two sets of fixed points. The first is the trivial fixed points at ${\bf s}_0 = \pm \left( 1, 0, 0 \right)$ which are stable fixed points for all parameters. If the system evolves towards these fixed points, the $4\pi$-periodic Josephson current in Eq.~(\ref{eq:EOM1}) vanishes and all experimental $4\pi$-periodic signatures would disappear. We also find another set of fixed points at\cite{supp}
\begin{equation}
{\bf s}_1 = \pm \left(E'_{\rm M}+{V_0 B_f}/{R}, 0, \sqrt{1-(E'_{\rm M}+{V_0 B_f}/{R})^2 } \right),
\end{equation}
which are stable fixed points only when the injected current is small so that the dc voltage is smaller than a critical value of $V_c = |E_{\rm M} \alpha - E'_{\rm M}|R/B_f$. The existence of these fixed points is voltage-dependent, which is qualitatively different from the trivial fixed points.

\begin{figure}[t]
\begin{center}
\includegraphics[clip = true, width = 0.9 \columnwidth]{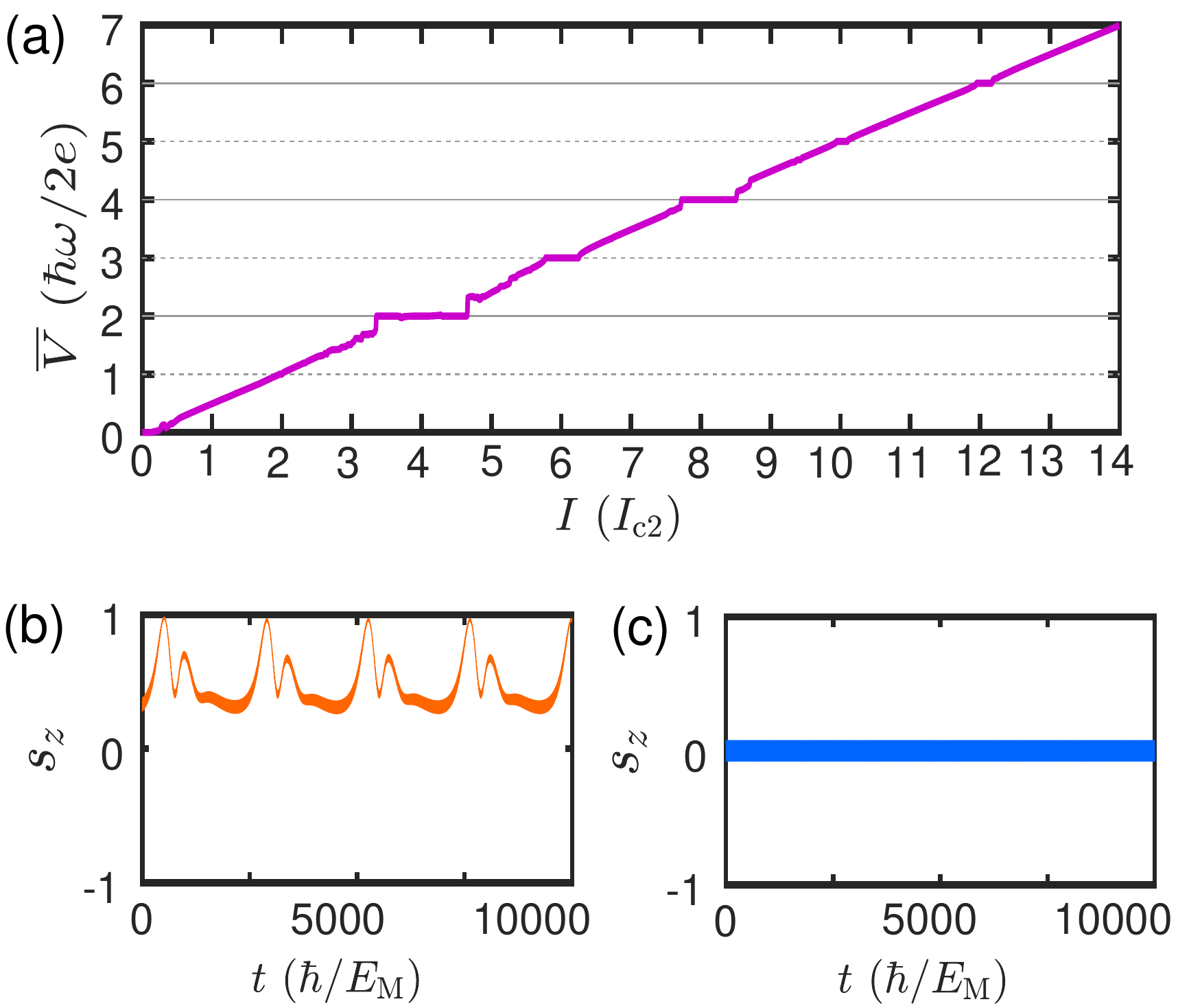}
\end{center}
\caption{(a) The Shapiro steps for the topological Josephson junction simulated with Eq. (\ref{eq:EOM}). The first Shapiro step is strongly suppressed while all other steps are clearly visible. 
(b) The time evolution of $s_z$ for the voltage around the first Shapiro step. The pseudo spin oscillates with a non-zero averaging value, and the resulted $4\pi$-periodic supercurrent strongly suppresses the first step.
(c) The time evolution of $s_z$ for the voltage around the third Shapiro step. The pseudo spin goes to the fixed point of $s_z \approx 0$ which effectively shuts down the $4\pi$-periodic channel for the Josephson current.}
\label{fig:shapiro}
\end{figure}

These analytical results for fixed points provide insight into the experimentally reported voltage-dependent behaviors of topological junctions\cite{oostinga2013,deacon2017radiation}. From the fixed point analysis, we find two different voltage regimes. At dc voltage below $V_c$, there are two sets of fixed points, and the system has a chance of evolving to either of them. If the system evolves to the fixed point ${\bf s}_1$ as shown in Fig. \ref{fig:FX}a, the final stable state would have a non-vanishing $s_z$ and therefore a non-vanishing $4\pi$-periodic Josephson current shows up in the equation for the Josephson phase. In this voltage regime, we should expect transport signatures for $4\pi$-periodicity.
However, for the voltage above $V_c$, there exits only the trivial fixed points at ${\bf s}_0$. When the system evolves towards it as shown in Fig. \ref{fig:FX}b, the final stable state would have a vanishing $s_z$, and the $4\pi$-periodic Josephson current vanishes. In this voltage regime, all the transport signatures for the $4\pi$-periodicity should disappear. 
Based on these fixed point analysis, we predict that the transport of topological junctions would exhibit nontrivial $4\pi$-periodic signatures only at low voltage, while at high voltages it would look quite similar to the trivial junctions.

{\it {\color{blue} Shapiro steps.}}-- The Shapiro steps are the plateaus of the I-V curve at voltages $V_n = n \hbar \omega/2e $ under an injected ac current with frequency $\omega$. It is a powerful tool for probing the dynamics of Josephson junctions since it reflects the resonance between the dc and the ac Josephson relation. For topological junctions, it was anticipated that the odd number Shapiro steps with $n=1,3,5\ldots$ should be suppressed by the $4\pi$-periodic supercurrent. 
The experimental results, however, often show strong suppression of low order odd number steps such as the one with $n=1$, while other odd number Shapiro steps at higher voltages are robust. Since the understanding of the experimental results are crucial for detecting Majorana zero modes, it is timely to implement the Eq.~(\ref{eq:EOM}) to calculate the Shapiro steps of Majorana Josephson junctions.

We consider 
an injected current of $I_{\rm ex}(t) = I + I' \cos \omega t$ and calculate the I-V curve of the junction, with the results for a typical junction parameter shown in Fig.~\ref{fig:shapiro}a. We find Shapiro steps at $V = n \hbar \omega/2e $, where $n$ labels the number of the step. Intriguingly, it is clear that the first Shapiro step with $n=1$ is strongly suppressed, while all other steps are clearly visible.
At first glance, the suppression of only one Shapiro step seems mysterious. One would expect a suppression of all odd-number steps if the $4\pi$-periodic supercurrent carried by the Majorana qubit is significant, or no suppression to any of the steps if the $4\pi$-periodic supercurrent is irrelevant.
For this phenomenon, our theory provides a possible mechanism: the feedback from the dynamics of the Majorana qubit. As we have shown in the analytical results, the Majorana qubit evolves to different stable states for different voltages. 
We examine the qubit dynamics at the voltages for the first step and the third step. As shown in Fig. \ref{fig:shapiro}b, at the voltage where the first step should appear,
the Majorana qubit evolves to the stable state with a finite $s_z$. Then the $4\pi$-periodic Josephson current will dominate and the Shapiro step is suppressed.
For the higher voltage of the third step, however, the Majorana qubit evolves to a stable state with $s_z \approx 0$, as shown in Fig. \ref{fig:shapiro}c. 
Then the $4\pi$-periodic supercurrent is blocked, and the junction would behaves similar to a conventional junction presenting Shapiro steps.
This feed back of the qubit dynamics provides a simple mechanism for the suppression of the first Shapiro step, and gives a possible explanation to one of the puzzles in the experimental findings of topological superconductors.

We emphasize that, while our theory is derived for topological junctions with Majorana qubit, it is actually valid for any junction with an embedded qubit that can be described by the low energy effective Hamiltonian Eq. (\ref{eq:JosephsonH}). One such example is the Josephson junction with quantum dots\cite{oriekhov2021}. In this sense, our calculation of Shapiro steps provide a signal for the feedback of embedded qubit, instead of a unique signature of Majorana zero modes.

{\it {\color{blue} Conclusion.}}-- In summary, we constructed a semiclassical theory for the topological Josephson junctions with an embedded Majorana qubit. We revealed nontrivial qubit dynamics such as the Landau-Zener transitions and the anisotropic Gilbert damping. We found that the feedback of the qubit dynamics strongly modifies the transport features of the junction. We applied the theory to study the Shapiro steps of the topological junctions and demonstrated the suppression of the first Shapiro step which agrees with recent experiments. We reveal that this phenomenon is due to the voltage-selective feedback from the dynamics of Majorana qubit.

\textit{Acknowledgments.---} We thank Zhongbo Yan, Peng Ye and Shuai Yin for valuable discussions. This work was supported by NSFC (Grant No. 12174453), NKRDPC-2017YFA0206203, 2017YFA0303302, 2018YFA0305603, and Guangdong Basic and Applied Basic Research Foundation (Grant No. 2019A1515011620). Z.H. is supported by the Robert A.Welch Foundation under Grant No. E-1146.

\bibliography{Feynman}

\begin{thebibliography}{60}%
\makeatletter
\providecommand \@ifxundefined [1]{%
 \@ifx{#1\undefined}
}%
\providecommand \@ifnum [1]{%
 \ifnum #1\expandafter \@firstoftwo
 \else \expandafter \@secondoftwo
 \fi
}%
\providecommand \@ifx [1]{%
 \ifx #1\expandafter \@firstoftwo
 \else \expandafter \@secondoftwo
 \fi
}%
\providecommand \natexlab [1]{#1}%
\providecommand \enquote  [1]{``#1''}%
\providecommand \bibnamefont  [1]{#1}%
\providecommand \bibfnamefont [1]{#1}%
\providecommand \citenamefont [1]{#1}%
\providecommand \href@noop [0]{\@secondoftwo}%
\providecommand \href [0]{\begingroup \@sanitize@url \@href}%
\providecommand \@href[1]{\@@startlink{#1}\@@href}%
\providecommand \@@href[1]{\endgroup#1\@@endlink}%
\providecommand \@sanitize@url [0]{\catcode `\\12\catcode `\$12\catcode
  `\&12\catcode `\#12\catcode `\^12\catcode `\_12\catcode `\%12\relax}%
\providecommand \@@startlink[1]{}%
\providecommand \@@endlink[0]{}%
\providecommand \url  [0]{\begingroup\@sanitize@url \@url }%
\providecommand \@url [1]{\endgroup\@href {#1}{\urlprefix }}%
\providecommand \urlprefix  [0]{URL }%
\providecommand \Eprint [0]{\href }%
\providecommand \doibase [0]{https://doi.org/}%
\providecommand \selectlanguage [0]{\@gobble}%
\providecommand \bibinfo  [0]{\@secondoftwo}%
\providecommand \bibfield  [0]{\@secondoftwo}%
\providecommand \translation [1]{[#1]}%
\providecommand \BibitemOpen [0]{}%
\providecommand \bibitemStop [0]{}%
\providecommand \bibitemNoStop [0]{.\EOS\space}%
\providecommand \EOS [0]{\spacefactor3000\relax}%
\providecommand \BibitemShut  [1]{\csname bibitem#1\endcsname}%
\let\auto@bib@innerbib\@empty
\bibitem [{\citenamefont {Devoret}\ and\ \citenamefont
  {Schoelkopf}(2013)}]{devoret2013review}%
  \BibitemOpen
  \bibfield  {author} {\bibinfo {author} {\bibfnamefont {M.~H.}\ \bibnamefont
  {Devoret}}\ and\ \bibinfo {author} {\bibfnamefont {R.~J.}\ \bibnamefont
  {Schoelkopf}},\ }\bibfield  {title} {\bibinfo {title} {Superconducting
  circuits for quantum information: An outlook},\ }\href
  {https://doi.org/10.1126/science.1231930} {\bibfield  {journal} {\bibinfo
  {journal} {Science}\ }\textbf {\bibinfo {volume} {339}},\ \bibinfo {pages}
  {1169} (\bibinfo {year} {2013})}\BibitemShut {NoStop}%
\bibitem [{\citenamefont {Martinis}\ \emph {et~al.}(2020)\citenamefont
  {Martinis}, \citenamefont {Devoret},\ and\ \citenamefont
  {Clarke}}]{martinis2020}%
  \BibitemOpen
  \bibfield  {author} {\bibinfo {author} {\bibfnamefont {J.~M.}\ \bibnamefont
  {Martinis}}, \bibinfo {author} {\bibfnamefont {M.~H.}\ \bibnamefont
  {Devoret}},\ and\ \bibinfo {author} {\bibfnamefont {J.}~\bibnamefont
  {Clarke}},\ }\bibfield  {title} {\bibinfo {title} {Quantum josephson junction
  circuits and the dawn of artificial atoms},\ }\href
  {https://doi.org/10.1038/s41567-020-0829-5} {\bibfield  {journal} {\bibinfo
  {journal} {Nature Physics}\ }\textbf {\bibinfo {volume} {16}},\ \bibinfo
  {pages} {234} (\bibinfo {year} {2020})}\BibitemShut {NoStop}%
\bibitem [{\citenamefont {Kjaergaard}\ \emph {et~al.}(2020)\citenamefont
  {Kjaergaard}, \citenamefont {Schwartz}, \citenamefont {Braumüller},
  \citenamefont {Krantz}, \citenamefont {Wang}, \citenamefont {Gustavsson},\
  and\ \citenamefont {Oliver}}]{kjaergaard2019}%
  \BibitemOpen
  \bibfield  {author} {\bibinfo {author} {\bibfnamefont {M.}~\bibnamefont
  {Kjaergaard}}, \bibinfo {author} {\bibfnamefont {M.~E.}\ \bibnamefont
  {Schwartz}}, \bibinfo {author} {\bibfnamefont {J.}~\bibnamefont
  {Braumüller}}, \bibinfo {author} {\bibfnamefont {P.}~\bibnamefont {Krantz}},
  \bibinfo {author} {\bibfnamefont {J.~I.-J.}\ \bibnamefont {Wang}}, \bibinfo
  {author} {\bibfnamefont {S.}~\bibnamefont {Gustavsson}},\ and\ \bibinfo
  {author} {\bibfnamefont {W.~D.}\ \bibnamefont {Oliver}},\ }\bibfield  {title}
  {\bibinfo {title} {Superconducting qubits: Current state of play},\ }\href
  {https://doi.org/10.1146/annurev-conmatphys-031119-050605} {\bibfield
  {journal} {\bibinfo  {journal} {Annual Review of Condensed Matter Physics}\
  }\textbf {\bibinfo {volume} {11}},\ \bibinfo {pages} {369} (\bibinfo {year}
  {2020})}\BibitemShut {NoStop}%
\bibitem [{\citenamefont {Siddiqi}(2021)}]{siddiqi2021}%
  \BibitemOpen
  \bibfield  {author} {\bibinfo {author} {\bibfnamefont {I.}~\bibnamefont
  {Siddiqi}},\ }\bibfield  {title} {\bibinfo {title} {Engineering
  high-coherence superconducting qubits},\ }\href
  {https://doi.org/10.1038/s41578-021-00370-4} {\bibfield  {journal} {\bibinfo
  {journal} {Nature Reviews Materials}\ }\textbf {\bibinfo {volume} {6}},\
  \bibinfo {pages} {875} (\bibinfo {year} {2021})}\BibitemShut {NoStop}%
\bibitem [{\citenamefont {Tinkham}(2004)}]{tinkham2004}%
  \BibitemOpen
  \bibfield  {author} {\bibinfo {author} {\bibfnamefont {M.}~\bibnamefont
  {Tinkham}},\ }\href@noop {} {\emph {\bibinfo {title} {Introduction to
  superconductivity}}}\ (\bibinfo  {publisher} {Courier Corporation},\ \bibinfo
  {year} {2004})\BibitemShut {NoStop}%
\bibitem [{\citenamefont {Feng}\ \emph {et~al.}(2018)\citenamefont {Feng},
  \citenamefont {Huang}, \citenamefont {Wang},\ and\ \citenamefont
  {Niu}}]{feng2018hysteresis}%
  \BibitemOpen
  \bibfield  {author} {\bibinfo {author} {\bibfnamefont {J.-J.}\ \bibnamefont
  {Feng}}, \bibinfo {author} {\bibfnamefont {Z.}~\bibnamefont {Huang}},
  \bibinfo {author} {\bibfnamefont {Z.}~\bibnamefont {Wang}},\ and\ \bibinfo
  {author} {\bibfnamefont {Q.}~\bibnamefont {Niu}},\ }\bibfield  {title}
  {\bibinfo {title} {Hysteresis from nonlinear dynamics of majorana modes in
  topological josephson junctions},\ }\href
  {https://doi.org/10.1103/PhysRevB.98.134515} {\bibfield  {journal} {\bibinfo
  {journal} {Phys. Rev. B}\ }\textbf {\bibinfo {volume} {98}},\ \bibinfo
  {pages} {134515} (\bibinfo {year} {2018})}\BibitemShut {NoStop}%
\bibitem [{\citenamefont {Choi}\ \emph {et~al.}(2020)\citenamefont {Choi},
  \citenamefont {Calzona},\ and\ \citenamefont {Trauzettel}}]{choi2020}%
  \BibitemOpen
  \bibfield  {author} {\bibinfo {author} {\bibfnamefont {S.-J.}\ \bibnamefont
  {Choi}}, \bibinfo {author} {\bibfnamefont {A.}~\bibnamefont {Calzona}},\ and\
  \bibinfo {author} {\bibfnamefont {B.}~\bibnamefont {Trauzettel}},\ }\bibfield
   {title} {\bibinfo {title} {Majorana-induced dc shapiro steps in topological
  josephson junctions},\ }\href {https://doi.org/10.1103/PhysRevB.102.140501}
  {\bibfield  {journal} {\bibinfo  {journal} {Phys. Rev. B}\ }\textbf {\bibinfo
  {volume} {102}},\ \bibinfo {pages} {140501} (\bibinfo {year}
  {2020})}\BibitemShut {NoStop}%
\bibitem [{\citenamefont {Oriekhov}\ \emph {et~al.}(2021)\citenamefont
  {Oriekhov}, \citenamefont {Cheipesh},\ and\ \citenamefont
  {Beenakker}}]{oriekhov2021}%
  \BibitemOpen
  \bibfield  {author} {\bibinfo {author} {\bibfnamefont {D.~O.}\ \bibnamefont
  {Oriekhov}}, \bibinfo {author} {\bibfnamefont {Y.}~\bibnamefont {Cheipesh}},\
  and\ \bibinfo {author} {\bibfnamefont {C.~W.~J.}\ \bibnamefont {Beenakker}},\
  }\bibfield  {title} {\bibinfo {title} {Voltage staircase in a current-biased
  quantum-dot josephson junction},\ }\href
  {https://doi.org/10.1103/PhysRevB.103.094518} {\bibfield  {journal} {\bibinfo
   {journal} {Phys. Rev. B}\ }\textbf {\bibinfo {volume} {103}},\ \bibinfo
  {pages} {094518} (\bibinfo {year} {2021})}\BibitemShut {NoStop}%
\bibitem [{\citenamefont {Deng}\ \emph {et~al.}(2012)\citenamefont {Deng},
  \citenamefont {Yu}, \citenamefont {Huang}, \citenamefont {Larsson},
  \citenamefont {Caroff},\ and\ \citenamefont {Xu}}]{deng2012anomalous}%
  \BibitemOpen
  \bibfield  {author} {\bibinfo {author} {\bibfnamefont {M.~T.}\ \bibnamefont
  {Deng}}, \bibinfo {author} {\bibfnamefont {C.~L.}\ \bibnamefont {Yu}},
  \bibinfo {author} {\bibfnamefont {G.~Y.}\ \bibnamefont {Huang}}, \bibinfo
  {author} {\bibfnamefont {M.}~\bibnamefont {Larsson}}, \bibinfo {author}
  {\bibfnamefont {P.}~\bibnamefont {Caroff}},\ and\ \bibinfo {author}
  {\bibfnamefont {H.~Q.}\ \bibnamefont {Xu}},\ }\bibfield  {title} {\bibinfo
  {title} {Anomalous zero-bias conductance peak in a nb–insb nanowire–nb
  hybrid device},\ }\href {https://doi.org/10.1021/nl303758w} {\bibfield
  {journal} {\bibinfo  {journal} {Nano Letters}\ }\textbf {\bibinfo {volume}
  {12}},\ \bibinfo {pages} {6414} (\bibinfo {year} {2012})}\BibitemShut
  {NoStop}%
\bibitem [{\citenamefont {Oostinga}\ \emph {et~al.}(2013)\citenamefont
  {Oostinga}, \citenamefont {Maier}, \citenamefont {Sch\"uffelgen},
  \citenamefont {Knott}, \citenamefont {Ames}, \citenamefont {Br\"une},
  \citenamefont {Tkachov}, \citenamefont {Buhmann},\ and\ \citenamefont
  {Molenkamp}}]{oostinga2013}%
  \BibitemOpen
  \bibfield  {author} {\bibinfo {author} {\bibfnamefont {J.~B.}\ \bibnamefont
  {Oostinga}}, \bibinfo {author} {\bibfnamefont {L.}~\bibnamefont {Maier}},
  \bibinfo {author} {\bibfnamefont {P.}~\bibnamefont {Sch\"uffelgen}}, \bibinfo
  {author} {\bibfnamefont {D.}~\bibnamefont {Knott}}, \bibinfo {author}
  {\bibfnamefont {C.}~\bibnamefont {Ames}}, \bibinfo {author} {\bibfnamefont
  {C.}~\bibnamefont {Br\"une}}, \bibinfo {author} {\bibfnamefont
  {G.}~\bibnamefont {Tkachov}}, \bibinfo {author} {\bibfnamefont
  {H.}~\bibnamefont {Buhmann}},\ and\ \bibinfo {author} {\bibfnamefont {L.~W.}\
  \bibnamefont {Molenkamp}},\ }\bibfield  {title} {\bibinfo {title} {Josephson
  supercurrent through the topological surface states of strained bulk hgte},\
  }\href {https://doi.org/10.1103/PhysRevX.3.021007} {\bibfield  {journal}
  {\bibinfo  {journal} {Phys. Rev. X}\ }\textbf {\bibinfo {volume} {3}},\
  \bibinfo {pages} {021007} (\bibinfo {year} {2013})}\BibitemShut {NoStop}%
\bibitem [{\citenamefont {Peng}\ \emph {et~al.}(2016)\citenamefont {Peng},
  \citenamefont {Vinkler-Aviv}, \citenamefont {Brouwer}, \citenamefont
  {Glazman},\ and\ \citenamefont {von Oppen}}]{peng2016TJJ}%
  \BibitemOpen
  \bibfield  {author} {\bibinfo {author} {\bibfnamefont {Y.}~\bibnamefont
  {Peng}}, \bibinfo {author} {\bibfnamefont {Y.}~\bibnamefont {Vinkler-Aviv}},
  \bibinfo {author} {\bibfnamefont {P.~W.}\ \bibnamefont {Brouwer}}, \bibinfo
  {author} {\bibfnamefont {L.~I.}\ \bibnamefont {Glazman}},\ and\ \bibinfo
  {author} {\bibfnamefont {F.}~\bibnamefont {von Oppen}},\ }\bibfield  {title}
  {\bibinfo {title} {Parity anomaly and spin transmutation in quantum spin hall
  josephson junctions},\ }\href
  {https://doi.org/10.1103/PhysRevLett.117.267001} {\bibfield  {journal}
  {\bibinfo  {journal} {Phys. Rev. Lett.}\ }\textbf {\bibinfo {volume} {117}},\
  \bibinfo {pages} {267001} (\bibinfo {year} {2016})}\BibitemShut {NoStop}%
\bibitem [{\citenamefont {Cayao}\ \emph {et~al.}(2017)\citenamefont {Cayao},
  \citenamefont {San-Jose}, \citenamefont {Black-Schaffer}, \citenamefont
  {Aguado},\ and\ \citenamefont {Prada}}]{cayao2017}%
  \BibitemOpen
  \bibfield  {author} {\bibinfo {author} {\bibfnamefont {J.}~\bibnamefont
  {Cayao}}, \bibinfo {author} {\bibfnamefont {P.}~\bibnamefont {San-Jose}},
  \bibinfo {author} {\bibfnamefont {A.~M.}\ \bibnamefont {Black-Schaffer}},
  \bibinfo {author} {\bibfnamefont {R.}~\bibnamefont {Aguado}},\ and\ \bibinfo
  {author} {\bibfnamefont {E.}~\bibnamefont {Prada}},\ }\bibfield  {title}
  {\bibinfo {title} {Majorana splitting from critical currents in josephson
  junctions},\ }\href {https://doi.org/10.1103/PhysRevB.96.205425} {\bibfield
  {journal} {\bibinfo  {journal} {Phys. Rev. B}\ }\textbf {\bibinfo {volume}
  {96}},\ \bibinfo {pages} {205425} (\bibinfo {year} {2017})}\BibitemShut
  {NoStop}%
\bibitem [{\citenamefont {Deacon}\ \emph {et~al.}(2017)\citenamefont {Deacon},
  \citenamefont {Wiedenmann}, \citenamefont {Bocquillon}, \citenamefont
  {Dom\'{\i}nguez}, \citenamefont {Klapwijk}, \citenamefont {Leubner},
  \citenamefont {Br\"une}, \citenamefont {Hankiewicz}, \citenamefont {Tarucha},
  \citenamefont {Ishibashi}, \citenamefont {Buhmann},\ and\ \citenamefont
  {Molenkamp}}]{deacon2017radiation}%
  \BibitemOpen
  \bibfield  {author} {\bibinfo {author} {\bibfnamefont {R.~S.}\ \bibnamefont
  {Deacon}}, \bibinfo {author} {\bibfnamefont {J.}~\bibnamefont {Wiedenmann}},
  \bibinfo {author} {\bibfnamefont {E.}~\bibnamefont {Bocquillon}}, \bibinfo
  {author} {\bibfnamefont {F.}~\bibnamefont {Dom\'{\i}nguez}}, \bibinfo
  {author} {\bibfnamefont {T.~M.}\ \bibnamefont {Klapwijk}}, \bibinfo {author}
  {\bibfnamefont {P.}~\bibnamefont {Leubner}}, \bibinfo {author} {\bibfnamefont
  {C.}~\bibnamefont {Br\"une}}, \bibinfo {author} {\bibfnamefont {E.~M.}\
  \bibnamefont {Hankiewicz}}, \bibinfo {author} {\bibfnamefont
  {S.}~\bibnamefont {Tarucha}}, \bibinfo {author} {\bibfnamefont
  {K.}~\bibnamefont {Ishibashi}}, \bibinfo {author} {\bibfnamefont
  {H.}~\bibnamefont {Buhmann}},\ and\ \bibinfo {author} {\bibfnamefont {L.~W.}\
  \bibnamefont {Molenkamp}},\ }\bibfield  {title} {\bibinfo {title} {Josephson
  radiation from gapless andreev bound states in hgte-based topological
  junctions},\ }\href {https://doi.org/10.1103/PhysRevX.7.021011} {\bibfield
  {journal} {\bibinfo  {journal} {Phys. Rev. X}\ }\textbf {\bibinfo {volume}
  {7}},\ \bibinfo {pages} {021011} (\bibinfo {year} {2017})}\BibitemShut
  {NoStop}%
\bibitem [{\citenamefont {Kamata}\ \emph {et~al.}(2018)\citenamefont {Kamata},
  \citenamefont {Deacon}, \citenamefont {Matsuo}, \citenamefont {Li},
  \citenamefont {Jeppesen}, \citenamefont {Samuelson}, \citenamefont {Xu},
  \citenamefont {Ishibashi},\ and\ \citenamefont {Tarucha}}]{kamata2018}%
  \BibitemOpen
  \bibfield  {author} {\bibinfo {author} {\bibfnamefont {H.}~\bibnamefont
  {Kamata}}, \bibinfo {author} {\bibfnamefont {R.~S.}\ \bibnamefont {Deacon}},
  \bibinfo {author} {\bibfnamefont {S.}~\bibnamefont {Matsuo}}, \bibinfo
  {author} {\bibfnamefont {K.}~\bibnamefont {Li}}, \bibinfo {author}
  {\bibfnamefont {S.}~\bibnamefont {Jeppesen}}, \bibinfo {author}
  {\bibfnamefont {L.}~\bibnamefont {Samuelson}}, \bibinfo {author}
  {\bibfnamefont {H.~Q.}\ \bibnamefont {Xu}}, \bibinfo {author} {\bibfnamefont
  {K.}~\bibnamefont {Ishibashi}},\ and\ \bibinfo {author} {\bibfnamefont
  {S.}~\bibnamefont {Tarucha}},\ }\bibfield  {title} {\bibinfo {title}
  {Anomalous modulation of josephson radiation in nanowire-based josephson
  junctions},\ }\href {https://doi.org/10.1103/PhysRevB.98.041302} {\bibfield
  {journal} {\bibinfo  {journal} {Phys. Rev. B}\ }\textbf {\bibinfo {volume}
  {98}},\ \bibinfo {pages} {041302} (\bibinfo {year} {2018})}\BibitemShut
  {NoStop}%
\bibitem [{\citenamefont {Schrade}\ and\ \citenamefont
  {Fu}(2018)}]{schrade2018}%
  \BibitemOpen
  \bibfield  {author} {\bibinfo {author} {\bibfnamefont {C.}~\bibnamefont
  {Schrade}}\ and\ \bibinfo {author} {\bibfnamefont {L.}~\bibnamefont {Fu}},\
  }\bibfield  {title} {\bibinfo {title} {Parity-controlled $2\ensuremath{\pi}$
  josephson effect mediated by majorana kramers pairs},\ }\href
  {https://doi.org/10.1103/PhysRevLett.120.267002} {\bibfield  {journal}
  {\bibinfo  {journal} {Phys. Rev. Lett.}\ }\textbf {\bibinfo {volume} {120}},\
  \bibinfo {pages} {267002} (\bibinfo {year} {2018})}\BibitemShut {NoStop}%
\bibitem [{\citenamefont {Lei}\ \emph {et~al.}(2018)\citenamefont {Lei},
  \citenamefont {Chen},\ and\ \citenamefont {MacDonald}}]{lei2018prl}%
  \BibitemOpen
  \bibfield  {author} {\bibinfo {author} {\bibfnamefont {C.}~\bibnamefont
  {Lei}}, \bibinfo {author} {\bibfnamefont {H.}~\bibnamefont {Chen}},\ and\
  \bibinfo {author} {\bibfnamefont {A.~H.}\ \bibnamefont {MacDonald}},\
  }\bibfield  {title} {\bibinfo {title} {Ultrathin films of superconducting
  metals as a platform for topological superconductivity},\ }\href
  {https://doi.org/10.1103/PhysRevLett.121.227701} {\bibfield  {journal}
  {\bibinfo  {journal} {Phys. Rev. Lett.}\ }\textbf {\bibinfo {volume} {121}},\
  \bibinfo {pages} {227701} (\bibinfo {year} {2018})}\BibitemShut {NoStop}%
\bibitem [{\citenamefont {Li}\ \emph {et~al.}(2018{\natexlab{a}})\citenamefont
  {Li}, \citenamefont {Song}, \citenamefont {Liu}, \citenamefont {Jiang},
  \citenamefont {Sun},\ and\ \citenamefont {Xie}}]{li2018prb}%
  \BibitemOpen
  \bibfield  {author} {\bibinfo {author} {\bibfnamefont {Y.-H.}\ \bibnamefont
  {Li}}, \bibinfo {author} {\bibfnamefont {J.}~\bibnamefont {Song}}, \bibinfo
  {author} {\bibfnamefont {J.}~\bibnamefont {Liu}}, \bibinfo {author}
  {\bibfnamefont {H.}~\bibnamefont {Jiang}}, \bibinfo {author} {\bibfnamefont
  {Q.-F.}\ \bibnamefont {Sun}},\ and\ \bibinfo {author} {\bibfnamefont {X.~C.}\
  \bibnamefont {Xie}},\ }\bibfield  {title} {\bibinfo {title} {Doubled shapiro
  steps in a topological josephson junction},\ }\href
  {https://doi.org/10.1103/PhysRevB.97.045423} {\bibfield  {journal} {\bibinfo
  {journal} {Phys. Rev. B}\ }\textbf {\bibinfo {volume} {97}},\ \bibinfo
  {pages} {045423} (\bibinfo {year} {2018}{\natexlab{a}})}\BibitemShut
  {NoStop}%
\bibitem [{\citenamefont {Liu}\ \emph {et~al.}(2019)\citenamefont {Liu},
  \citenamefont {Deng},\ and\ \citenamefont {Wakabayashi}}]{liu2019PRL}%
  \BibitemOpen
  \bibfield  {author} {\bibinfo {author} {\bibfnamefont {F.}~\bibnamefont
  {Liu}}, \bibinfo {author} {\bibfnamefont {H.-Y.}\ \bibnamefont {Deng}},\ and\
  \bibinfo {author} {\bibfnamefont {K.}~\bibnamefont {Wakabayashi}},\
  }\bibfield  {title} {\bibinfo {title} {Helical topological edge states in a
  quadrupole phase},\ }\href {https://doi.org/10.1103/PhysRevLett.122.086804}
  {\bibfield  {journal} {\bibinfo  {journal} {Phys. Rev. Lett.}\ }\textbf
  {\bibinfo {volume} {122}},\ \bibinfo {pages} {086804} (\bibinfo {year}
  {2019})}\BibitemShut {NoStop}%
\bibitem [{\citenamefont {Laroche}\ \emph {et~al.}(2019)\citenamefont
  {Laroche}, \citenamefont {Bouman}, \citenamefont {van Woerkom}, \citenamefont
  {Proutski}, \citenamefont {Murthy}, \citenamefont {Pikulin}, \citenamefont
  {Nayak}, \citenamefont {van Gulik}, \citenamefont {Nyg{\aa}rd}, \citenamefont
  {Krogstrup}, \citenamefont {Kouwenhoven},\ and\ \citenamefont
  {Geresdi}}]{laroche2019radiation}%
  \BibitemOpen
  \bibfield  {author} {\bibinfo {author} {\bibfnamefont {D.}~\bibnamefont
  {Laroche}}, \bibinfo {author} {\bibfnamefont {D.}~\bibnamefont {Bouman}},
  \bibinfo {author} {\bibfnamefont {D.~J.}\ \bibnamefont {van Woerkom}},
  \bibinfo {author} {\bibfnamefont {A.}~\bibnamefont {Proutski}}, \bibinfo
  {author} {\bibfnamefont {C.}~\bibnamefont {Murthy}}, \bibinfo {author}
  {\bibfnamefont {D.~I.}\ \bibnamefont {Pikulin}}, \bibinfo {author}
  {\bibfnamefont {C.}~\bibnamefont {Nayak}}, \bibinfo {author} {\bibfnamefont
  {R.~J.~J.}\ \bibnamefont {van Gulik}}, \bibinfo {author} {\bibfnamefont
  {J.}~\bibnamefont {Nyg{\aa}rd}}, \bibinfo {author} {\bibfnamefont
  {P.}~\bibnamefont {Krogstrup}}, \bibinfo {author} {\bibfnamefont {L.~P.}\
  \bibnamefont {Kouwenhoven}},\ and\ \bibinfo {author} {\bibfnamefont
  {A.}~\bibnamefont {Geresdi}},\ }\bibfield  {title} {\bibinfo {title}
  {Observation of the 4p-periodic josephson effect in indium arsenide
  nanowires},\ }\href {https://doi.org/10.1038/s41467-018-08161-2} {\bibfield
  {journal} {\bibinfo  {journal} {Nature Communications}\ }\textbf {\bibinfo
  {volume} {10}},\ \bibinfo {pages} {245} (\bibinfo {year} {2019})}\BibitemShut
  {NoStop}%
\bibitem [{\citenamefont {Ren}\ \emph {et~al.}(2019)\citenamefont {Ren},
  \citenamefont {Pientka}, \citenamefont {Hart}, \citenamefont {Pierce},
  \citenamefont {Kosowsky}, \citenamefont {Lunczer}, \citenamefont {Schlereth},
  \citenamefont {Scharf}, \citenamefont {Hankiewicz}, \citenamefont
  {Molenkamp}, \citenamefont {Halperin},\ and\ \citenamefont
  {Yacoby}}]{ren2019josephson}%
  \BibitemOpen
  \bibfield  {author} {\bibinfo {author} {\bibfnamefont {H.}~\bibnamefont
  {Ren}}, \bibinfo {author} {\bibfnamefont {F.}~\bibnamefont {Pientka}},
  \bibinfo {author} {\bibfnamefont {S.}~\bibnamefont {Hart}}, \bibinfo {author}
  {\bibfnamefont {A.~T.}\ \bibnamefont {Pierce}}, \bibinfo {author}
  {\bibfnamefont {M.}~\bibnamefont {Kosowsky}}, \bibinfo {author}
  {\bibfnamefont {L.}~\bibnamefont {Lunczer}}, \bibinfo {author} {\bibfnamefont
  {R.}~\bibnamefont {Schlereth}}, \bibinfo {author} {\bibfnamefont
  {B.}~\bibnamefont {Scharf}}, \bibinfo {author} {\bibfnamefont {E.~M.}\
  \bibnamefont {Hankiewicz}}, \bibinfo {author} {\bibfnamefont {L.~W.}\
  \bibnamefont {Molenkamp}}, \bibinfo {author} {\bibfnamefont {B.~I.}\
  \bibnamefont {Halperin}},\ and\ \bibinfo {author} {\bibfnamefont
  {A.}~\bibnamefont {Yacoby}},\ }\bibfield  {title} {\bibinfo {title}
  {Topological superconductivity in a phase-controlled josephson junction},\
  }\href {https://doi.org/10.1038/s41586-019-1148-9} {\bibfield  {journal}
  {\bibinfo  {journal} {Nature}\ }\textbf {\bibinfo {volume} {569}},\ \bibinfo
  {pages} {93} (\bibinfo {year} {2019})}\BibitemShut {NoStop}%
\bibitem [{\citenamefont {Fornieri}\ \emph {et~al.}(2019)\citenamefont
  {Fornieri}, \citenamefont {Whiticar}, \citenamefont {Setiawan}, \citenamefont
  {Portol{\'e}s}, \citenamefont {Drachmann}, \citenamefont {Keselman},
  \citenamefont {Gronin}, \citenamefont {Thomas}, \citenamefont {Wang},
  \citenamefont {Kallaher}, \citenamefont {Gardner}, \citenamefont {Berg},
  \citenamefont {Manfra}, \citenamefont {Stern}, \citenamefont {Marcus},\ and\
  \citenamefont {Nichele}}]{fornieri2019}%
  \BibitemOpen
  \bibfield  {author} {\bibinfo {author} {\bibfnamefont {A.}~\bibnamefont
  {Fornieri}}, \bibinfo {author} {\bibfnamefont {A.~M.}\ \bibnamefont
  {Whiticar}}, \bibinfo {author} {\bibfnamefont {F.}~\bibnamefont {Setiawan}},
  \bibinfo {author} {\bibfnamefont {E.}~\bibnamefont {Portol{\'e}s}}, \bibinfo
  {author} {\bibfnamefont {A.~C.~C.}\ \bibnamefont {Drachmann}}, \bibinfo
  {author} {\bibfnamefont {A.}~\bibnamefont {Keselman}}, \bibinfo {author}
  {\bibfnamefont {S.}~\bibnamefont {Gronin}}, \bibinfo {author} {\bibfnamefont
  {C.}~\bibnamefont {Thomas}}, \bibinfo {author} {\bibfnamefont
  {T.}~\bibnamefont {Wang}}, \bibinfo {author} {\bibfnamefont {R.}~\bibnamefont
  {Kallaher}}, \bibinfo {author} {\bibfnamefont {G.~C.}\ \bibnamefont
  {Gardner}}, \bibinfo {author} {\bibfnamefont {E.}~\bibnamefont {Berg}},
  \bibinfo {author} {\bibfnamefont {M.~J.}\ \bibnamefont {Manfra}}, \bibinfo
  {author} {\bibfnamefont {A.}~\bibnamefont {Stern}}, \bibinfo {author}
  {\bibfnamefont {C.~M.}\ \bibnamefont {Marcus}},\ and\ \bibinfo {author}
  {\bibfnamefont {F.}~\bibnamefont {Nichele}},\ }\bibfield  {title} {\bibinfo
  {title} {Evidence of topological superconductivity in planar josephson
  junctions},\ }\href {https://doi.org/10.1038/s41586-019-1068-8} {\bibfield
  {journal} {\bibinfo  {journal} {Nature}\ }\textbf {\bibinfo {volume} {569}},\
  \bibinfo {pages} {89} (\bibinfo {year} {2019})}\BibitemShut {NoStop}%
\bibitem [{\citenamefont {He}\ \emph {et~al.}(2019)\citenamefont {He},
  \citenamefont {Liang}, \citenamefont {Tanaka},\ and\ \citenamefont
  {Nagaosa}}]{he2019}%
  \BibitemOpen
  \bibfield  {author} {\bibinfo {author} {\bibfnamefont {J.~J.}\ \bibnamefont
  {He}}, \bibinfo {author} {\bibfnamefont {T.}~\bibnamefont {Liang}}, \bibinfo
  {author} {\bibfnamefont {Y.}~\bibnamefont {Tanaka}},\ and\ \bibinfo {author}
  {\bibfnamefont {N.}~\bibnamefont {Nagaosa}},\ }\bibfield  {title} {\bibinfo
  {title} {Platform of chiral majorana edge modes and its quantum transport
  phenomena},\ }\href {https://doi.org/10.1038/s42005-019-0250-5} {\bibfield
  {journal} {\bibinfo  {journal} {Communications Physics}\ }\textbf {\bibinfo
  {volume} {2}},\ \bibinfo {pages} {149} (\bibinfo {year} {2019})}\BibitemShut
  {NoStop}%
\bibitem [{\citenamefont {Stern}\ and\ \citenamefont {Berg}(2019)}]{stern2019}%
  \BibitemOpen
  \bibfield  {author} {\bibinfo {author} {\bibfnamefont {A.}~\bibnamefont
  {Stern}}\ and\ \bibinfo {author} {\bibfnamefont {E.}~\bibnamefont {Berg}},\
  }\bibfield  {title} {\bibinfo {title} {Fractional josephson vortices and
  braiding of majorana zero modes in planar superconductor-semiconductor
  heterostructures},\ }\href {https://doi.org/10.1103/PhysRevLett.122.107701}
  {\bibfield  {journal} {\bibinfo  {journal} {Phys. Rev. Lett.}\ }\textbf
  {\bibinfo {volume} {122}},\ \bibinfo {pages} {107701} (\bibinfo {year}
  {2019})}\BibitemShut {NoStop}%
\bibitem [{\citenamefont {Klees}\ \emph {et~al.}(2020)\citenamefont {Klees},
  \citenamefont {Rastelli}, \citenamefont {Cuevas},\ and\ \citenamefont
  {Belzig}}]{klees2020}%
  \BibitemOpen
  \bibfield  {author} {\bibinfo {author} {\bibfnamefont {R.~L.}\ \bibnamefont
  {Klees}}, \bibinfo {author} {\bibfnamefont {G.}~\bibnamefont {Rastelli}},
  \bibinfo {author} {\bibfnamefont {J.~C.}\ \bibnamefont {Cuevas}},\ and\
  \bibinfo {author} {\bibfnamefont {W.}~\bibnamefont {Belzig}},\ }\bibfield
  {title} {\bibinfo {title} {Microwave spectroscopy reveals the quantum
  geometric tensor of topological josephson matter},\ }\href
  {https://doi.org/10.1103/PhysRevLett.124.197002} {\bibfield  {journal}
  {\bibinfo  {journal} {Phys. Rev. Lett.}\ }\textbf {\bibinfo {volume} {124}},\
  \bibinfo {pages} {197002} (\bibinfo {year} {2020})}\BibitemShut {NoStop}%
\bibitem [{\citenamefont {\'Avila}\ \emph {et~al.}(2020)\citenamefont
  {\'Avila}, \citenamefont {Prada}, \citenamefont {San-Jose},\ and\
  \citenamefont {Aguado}}]{avila2020}%
  \BibitemOpen
  \bibfield  {author} {\bibinfo {author} {\bibfnamefont {J.}~\bibnamefont
  {\'Avila}}, \bibinfo {author} {\bibfnamefont {E.}~\bibnamefont {Prada}},
  \bibinfo {author} {\bibfnamefont {P.}~\bibnamefont {San-Jose}},\ and\
  \bibinfo {author} {\bibfnamefont {R.}~\bibnamefont {Aguado}},\ }\bibfield
  {title} {\bibinfo {title} {Superconducting islands with topological josephson
  junctions based on semiconductor nanowires},\ }\href
  {https://doi.org/10.1103/PhysRevB.102.094518} {\bibfield  {journal} {\bibinfo
   {journal} {Phys. Rev. B}\ }\textbf {\bibinfo {volume} {102}},\ \bibinfo
  {pages} {094518} (\bibinfo {year} {2020})}\BibitemShut {NoStop}%
\bibitem [{\citenamefont {Razmadze}\ \emph {et~al.}(2020)\citenamefont
  {Razmadze}, \citenamefont {O'Farrell}, \citenamefont {Krogstrup},\ and\
  \citenamefont {Marcus}}]{razmade2020}%
  \BibitemOpen
  \bibfield  {author} {\bibinfo {author} {\bibfnamefont {D.}~\bibnamefont
  {Razmadze}}, \bibinfo {author} {\bibfnamefont {E.~C.~T.}\ \bibnamefont
  {O'Farrell}}, \bibinfo {author} {\bibfnamefont {P.}~\bibnamefont
  {Krogstrup}},\ and\ \bibinfo {author} {\bibfnamefont {C.~M.}\ \bibnamefont
  {Marcus}},\ }\bibfield  {title} {\bibinfo {title} {Quantum dot parity effects
  in trivial and topological josephson junctions},\ }\href
  {https://doi.org/10.1103/PhysRevLett.125.116803} {\bibfield  {journal}
  {\bibinfo  {journal} {Phys. Rev. Lett.}\ }\textbf {\bibinfo {volume} {125}},\
  \bibinfo {pages} {116803} (\bibinfo {year} {2020})}\BibitemShut {NoStop}%
\bibitem [{\citenamefont {Scharf}\ \emph {et~al.}(2021)\citenamefont {Scharf},
  \citenamefont {Braggio}, \citenamefont {Strambini}, \citenamefont
  {Giazotto},\ and\ \citenamefont {Hankiewicz}}]{scharf2021}%
  \BibitemOpen
  \bibfield  {author} {\bibinfo {author} {\bibfnamefont {B.}~\bibnamefont
  {Scharf}}, \bibinfo {author} {\bibfnamefont {A.}~\bibnamefont {Braggio}},
  \bibinfo {author} {\bibfnamefont {E.}~\bibnamefont {Strambini}}, \bibinfo
  {author} {\bibfnamefont {F.}~\bibnamefont {Giazotto}},\ and\ \bibinfo
  {author} {\bibfnamefont {E.~M.}\ \bibnamefont {Hankiewicz}},\ }\bibfield
  {title} {\bibinfo {title} {Thermodynamics in topological josephson
  junctions},\ }\href {https://doi.org/10.1103/PhysRevResearch.3.033062}
  {\bibfield  {journal} {\bibinfo  {journal} {Phys. Rev. Research}\ }\textbf
  {\bibinfo {volume} {3}},\ \bibinfo {pages} {033062} (\bibinfo {year}
  {2021})}\BibitemShut {NoStop}%
\bibitem [{\citenamefont {Dartiailh}\ \emph {et~al.}(2021)\citenamefont
  {Dartiailh}, \citenamefont {Mayer}, \citenamefont {Yuan}, \citenamefont
  {Wickramasinghe}, \citenamefont {Matos-Abiague}, \citenamefont {\ifmmode
  \check{Z}\else \v{Z}\fi{}uti\ifmmode~\acute{c}\else \'{c}\fi{}},\ and\
  \citenamefont {Shabani}}]{dartiailh2021}%
  \BibitemOpen
  \bibfield  {author} {\bibinfo {author} {\bibfnamefont {M.~C.}\ \bibnamefont
  {Dartiailh}}, \bibinfo {author} {\bibfnamefont {W.}~\bibnamefont {Mayer}},
  \bibinfo {author} {\bibfnamefont {J.}~\bibnamefont {Yuan}}, \bibinfo {author}
  {\bibfnamefont {K.~S.}\ \bibnamefont {Wickramasinghe}}, \bibinfo {author}
  {\bibfnamefont {A.}~\bibnamefont {Matos-Abiague}}, \bibinfo {author}
  {\bibfnamefont {I.}~\bibnamefont {\ifmmode \check{Z}\else
  \v{Z}\fi{}uti\ifmmode~\acute{c}\else \'{c}\fi{}}},\ and\ \bibinfo {author}
  {\bibfnamefont {J.}~\bibnamefont {Shabani}},\ }\bibfield  {title} {\bibinfo
  {title} {Phase signature of topological transition in josephson junctions},\
  }\href {https://doi.org/10.1103/PhysRevLett.126.036802} {\bibfield  {journal}
  {\bibinfo  {journal} {Phys. Rev. Lett.}\ }\textbf {\bibinfo {volume} {126}},\
  \bibinfo {pages} {036802} (\bibinfo {year} {2021})}\BibitemShut {NoStop}%
\bibitem [{\citenamefont {Li}\ \emph {et~al.}(2021)\citenamefont {Li},
  \citenamefont {Wang}, \citenamefont {Li}, \citenamefont {Zheng},
  \citenamefont {Brinkman}, \citenamefont {Yu},\ and\ \citenamefont
  {Liao}}]{li2021prl}%
  \BibitemOpen
  \bibfield  {author} {\bibinfo {author} {\bibfnamefont {C.-Z.}\ \bibnamefont
  {Li}}, \bibinfo {author} {\bibfnamefont {A.-Q.}\ \bibnamefont {Wang}},
  \bibinfo {author} {\bibfnamefont {C.}~\bibnamefont {Li}}, \bibinfo {author}
  {\bibfnamefont {W.-Z.}\ \bibnamefont {Zheng}}, \bibinfo {author}
  {\bibfnamefont {A.}~\bibnamefont {Brinkman}}, \bibinfo {author}
  {\bibfnamefont {D.-P.}\ \bibnamefont {Yu}},\ and\ \bibinfo {author}
  {\bibfnamefont {Z.-M.}\ \bibnamefont {Liao}},\ }\bibfield  {title} {\bibinfo
  {title} {Topological transition of superconductivity in dirac semimetal
  nanowire josephson junctions},\ }\href
  {https://doi.org/10.1103/PhysRevLett.126.027001} {\bibfield  {journal}
  {\bibinfo  {journal} {Phys. Rev. Lett.}\ }\textbf {\bibinfo {volume} {126}},\
  \bibinfo {pages} {027001} (\bibinfo {year} {2021})}\BibitemShut {NoStop}%
\bibitem [{\citenamefont {Jian}\ and\ \citenamefont {Yin}(2021)}]{jian2021}%
  \BibitemOpen
  \bibfield  {author} {\bibinfo {author} {\bibfnamefont {S.-K.}\ \bibnamefont
  {Jian}}\ and\ \bibinfo {author} {\bibfnamefont {S.}~\bibnamefont {Yin}},\
  }\bibfield  {title} {\bibinfo {title} {Chiral topological superconductivity
  in josephson junctions},\ }\href
  {https://doi.org/10.1103/PhysRevB.103.134514} {\bibfield  {journal} {\bibinfo
   {journal} {Phys. Rev. B}\ }\textbf {\bibinfo {volume} {103}},\ \bibinfo
  {pages} {134514} (\bibinfo {year} {2021})}\BibitemShut {NoStop}%
\bibitem [{\citenamefont {Zhang}\ and\ \citenamefont
  {Das~Sarma}(2021)}]{zhang2021prl}%
  \BibitemOpen
  \bibfield  {author} {\bibinfo {author} {\bibfnamefont {R.-X.}\ \bibnamefont
  {Zhang}}\ and\ \bibinfo {author} {\bibfnamefont {S.}~\bibnamefont
  {Das~Sarma}},\ }\bibfield  {title} {\bibinfo {title} {Anomalous floquet
  chiral topological superconductivity in a topological insulator sandwich
  structure},\ }\href {https://doi.org/10.1103/PhysRevLett.127.067001}
  {\bibfield  {journal} {\bibinfo  {journal} {Phys. Rev. Lett.}\ }\textbf
  {\bibinfo {volume} {127}},\ \bibinfo {pages} {067001} (\bibinfo {year}
  {2021})}\BibitemShut {NoStop}%
\bibitem [{\citenamefont {Kitaev}(2001)}]{kitaev2001unpaired}%
  \BibitemOpen
  \bibfield  {author} {\bibinfo {author} {\bibfnamefont {A.}~\bibnamefont
  {Kitaev}},\ }\bibfield  {title} {\bibinfo {title} {Unpaired majorana fermions
  in quantum wires},\ }\href {https://doi.org/10.1070/1063-7869/44/10S/S29}
  {\bibfield  {journal} {\bibinfo  {journal} {Physics-Uspekhi}\ }\textbf
  {\bibinfo {volume} {44}},\ \bibinfo {pages} {131} (\bibinfo {year}
  {2001})}\BibitemShut {NoStop}%
\bibitem [{\citenamefont {Kwon}\ \emph {et~al.}(2004)\citenamefont {Kwon},
  \citenamefont {Sengupta},\ and\ \citenamefont {Yakovenko}}]{Kwon2004}%
  \BibitemOpen
  \bibfield  {author} {\bibinfo {author} {\bibfnamefont {H.-J.}\ \bibnamefont
  {Kwon}}, \bibinfo {author} {\bibfnamefont {K.}~\bibnamefont {Sengupta}},\
  and\ \bibinfo {author} {\bibfnamefont {V.~M.}\ \bibnamefont {Yakovenko}},\
  }\bibfield  {title} {\bibinfo {title} {Fractional ac josephson effect in p-
  and d-wave superconductors},\ }\href
  {https://doi.org/10.1140/epjb/e2004-00066-4} {\bibfield  {journal} {\bibinfo
  {journal} {The European Physical Journal B - Condensed Matter and Complex
  Systems}\ }\textbf {\bibinfo {volume} {37}},\ \bibinfo {pages} {349}
  (\bibinfo {year} {2004})}\BibitemShut {NoStop}%
\bibitem [{\citenamefont {Dom\'{\i}nguez}\ \emph {et~al.}(2012)\citenamefont
  {Dom\'{\i}nguez}, \citenamefont {Hassler},\ and\ \citenamefont
  {Platero}}]{dominguez2012}%
  \BibitemOpen
  \bibfield  {author} {\bibinfo {author} {\bibfnamefont {F.}~\bibnamefont
  {Dom\'{\i}nguez}}, \bibinfo {author} {\bibfnamefont {F.}~\bibnamefont
  {Hassler}},\ and\ \bibinfo {author} {\bibfnamefont {G.}~\bibnamefont
  {Platero}},\ }\bibfield  {title} {\bibinfo {title} {Dynamical detection of
  majorana fermions in current-biased nanowires},\ }\href
  {https://doi.org/10.1103/PhysRevB.86.140503} {\bibfield  {journal} {\bibinfo
  {journal} {Phys. Rev. B}\ }\textbf {\bibinfo {volume} {86}},\ \bibinfo
  {pages} {140503} (\bibinfo {year} {2012})}\BibitemShut {NoStop}%
\bibitem [{\citenamefont {Dom\'{\i}nguez}\ \emph {et~al.}(2017)\citenamefont
  {Dom\'{\i}nguez}, \citenamefont {Kashuba}, \citenamefont {Bocquillon},
  \citenamefont {Wiedenmann}, \citenamefont {Deacon}, \citenamefont {Klapwijk},
  \citenamefont {Platero}, \citenamefont {Molenkamp}, \citenamefont
  {Trauzettel},\ and\ \citenamefont {Hankiewicz}}]{dominguez2017}%
  \BibitemOpen
  \bibfield  {author} {\bibinfo {author} {\bibfnamefont {F.}~\bibnamefont
  {Dom\'{\i}nguez}}, \bibinfo {author} {\bibfnamefont {O.}~\bibnamefont
  {Kashuba}}, \bibinfo {author} {\bibfnamefont {E.}~\bibnamefont {Bocquillon}},
  \bibinfo {author} {\bibfnamefont {J.}~\bibnamefont {Wiedenmann}}, \bibinfo
  {author} {\bibfnamefont {R.~S.}\ \bibnamefont {Deacon}}, \bibinfo {author}
  {\bibfnamefont {T.~M.}\ \bibnamefont {Klapwijk}}, \bibinfo {author}
  {\bibfnamefont {G.}~\bibnamefont {Platero}}, \bibinfo {author} {\bibfnamefont
  {L.~W.}\ \bibnamefont {Molenkamp}}, \bibinfo {author} {\bibfnamefont
  {B.}~\bibnamefont {Trauzettel}},\ and\ \bibinfo {author} {\bibfnamefont
  {E.~M.}\ \bibnamefont {Hankiewicz}},\ }\bibfield  {title} {\bibinfo {title}
  {Josephson junction dynamics in the presence of $2\ensuremath{\pi}$- and
  $4\ensuremath{\pi}$-periodic supercurrents},\ }\href
  {https://doi.org/10.1103/PhysRevB.95.195430} {\bibfield  {journal} {\bibinfo
  {journal} {Phys. Rev. B}\ }\textbf {\bibinfo {volume} {95}},\ \bibinfo
  {pages} {195430} (\bibinfo {year} {2017})}\BibitemShut {NoStop}%
\bibitem [{\citenamefont {Svetogorov}\ \emph {et~al.}(2020)\citenamefont
  {Svetogorov}, \citenamefont {Loss},\ and\ \citenamefont
  {Klinovaja}}]{svetogorov2020}%
  \BibitemOpen
  \bibfield  {author} {\bibinfo {author} {\bibfnamefont {A.~E.}\ \bibnamefont
  {Svetogorov}}, \bibinfo {author} {\bibfnamefont {D.}~\bibnamefont {Loss}},\
  and\ \bibinfo {author} {\bibfnamefont {J.}~\bibnamefont {Klinovaja}},\
  }\bibfield  {title} {\bibinfo {title} {Critical current for an insulating
  regime of an underdamped current-biased topological josephson junction},\
  }\href {https://doi.org/10.1103/PhysRevResearch.2.033448} {\bibfield
  {journal} {\bibinfo  {journal} {Phys. Rev. Research}\ }\textbf {\bibinfo
  {volume} {2}},\ \bibinfo {pages} {033448} (\bibinfo {year}
  {2020})}\BibitemShut {NoStop}%
\bibitem [{\citenamefont {Frombach}\ and\ \citenamefont
  {Recher}(2020)}]{frombach2020}%
  \BibitemOpen
  \bibfield  {author} {\bibinfo {author} {\bibfnamefont {D.}~\bibnamefont
  {Frombach}}\ and\ \bibinfo {author} {\bibfnamefont {P.}~\bibnamefont
  {Recher}},\ }\bibfield  {title} {\bibinfo {title} {Quasiparticle poisoning
  effects on the dynamics of topological josephson junctions},\ }\href
  {https://doi.org/10.1103/PhysRevB.101.115304} {\bibfield  {journal} {\bibinfo
   {journal} {Phys. Rev. B}\ }\textbf {\bibinfo {volume} {101}},\ \bibinfo
  {pages} {115304} (\bibinfo {year} {2020})}\BibitemShut {NoStop}%
\bibitem [{\citenamefont {Rokhinson}\ \emph {et~al.}(2012)\citenamefont
  {Rokhinson}, \citenamefont {Liu},\ and\ \citenamefont
  {Furdyna}}]{rokhinson2012fractional}%
  \BibitemOpen
  \bibfield  {author} {\bibinfo {author} {\bibfnamefont {L.~P.}\ \bibnamefont
  {Rokhinson}}, \bibinfo {author} {\bibfnamefont {X.}~\bibnamefont {Liu}},\
  and\ \bibinfo {author} {\bibfnamefont {J.~K.}\ \bibnamefont {Furdyna}},\
  }\bibfield  {title} {\bibinfo {title} {The fractional a.c. josephson effect
  in a semiconductor-superconductor nanowire as a signature of majorana
  particles},\ }\href {https://doi.org/10.1038/nphys2429} {\bibfield  {journal}
  {\bibinfo  {journal} {Nature Physics}\ }\textbf {\bibinfo {volume} {8}},\
  \bibinfo {pages} {795} (\bibinfo {year} {2012})}\BibitemShut {NoStop}%
\bibitem [{\citenamefont {Wiedenmann}\ \emph {et~al.}(2016)\citenamefont
  {Wiedenmann}, \citenamefont {Bocquillon}, \citenamefont {Deacon},
  \citenamefont {Hartinger}, \citenamefont {Herrmann}, \citenamefont
  {Klapwijk}, \citenamefont {Maier}, \citenamefont {Ames}, \citenamefont
  {Br{\"u}ne}, \citenamefont {Gould}, \citenamefont {Oiwa}, \citenamefont
  {Ishibashi}, \citenamefont {Tarucha}, \citenamefont {Buhmann},\ and\
  \citenamefont {Molenkamp}}]{wiedenmann2016shapiro4pi}%
  \BibitemOpen
  \bibfield  {author} {\bibinfo {author} {\bibfnamefont {J.}~\bibnamefont
  {Wiedenmann}}, \bibinfo {author} {\bibfnamefont {E.}~\bibnamefont
  {Bocquillon}}, \bibinfo {author} {\bibfnamefont {R.~S.}\ \bibnamefont
  {Deacon}}, \bibinfo {author} {\bibfnamefont {S.}~\bibnamefont {Hartinger}},
  \bibinfo {author} {\bibfnamefont {O.}~\bibnamefont {Herrmann}}, \bibinfo
  {author} {\bibfnamefont {T.~M.}\ \bibnamefont {Klapwijk}}, \bibinfo {author}
  {\bibfnamefont {L.}~\bibnamefont {Maier}}, \bibinfo {author} {\bibfnamefont
  {C.}~\bibnamefont {Ames}}, \bibinfo {author} {\bibfnamefont {C.}~\bibnamefont
  {Br{\"u}ne}}, \bibinfo {author} {\bibfnamefont {C.}~\bibnamefont {Gould}},
  \bibinfo {author} {\bibfnamefont {A.}~\bibnamefont {Oiwa}}, \bibinfo {author}
  {\bibfnamefont {K.}~\bibnamefont {Ishibashi}}, \bibinfo {author}
  {\bibfnamefont {S.}~\bibnamefont {Tarucha}}, \bibinfo {author} {\bibfnamefont
  {H.}~\bibnamefont {Buhmann}},\ and\ \bibinfo {author} {\bibfnamefont {L.~W.}\
  \bibnamefont {Molenkamp}},\ }\bibfield  {title} {\bibinfo {title}
  {4p-periodic josephson supercurrent in hgte-based topological josephson
  junctions},\ }\href {https://doi.org/10.1038/ncomms10303} {\bibfield
  {journal} {\bibinfo  {journal} {Nature Communications}\ }\textbf {\bibinfo
  {volume} {7}},\ \bibinfo {pages} {10303} (\bibinfo {year}
  {2016})}\BibitemShut {NoStop}%
\bibitem [{\citenamefont {Bocquillon}\ \emph {et~al.}(2017)\citenamefont
  {Bocquillon}, \citenamefont {Deacon}, \citenamefont {Wiedenmann},
  \citenamefont {Leubner}, \citenamefont {Klapwijk}, \citenamefont {Br{\"u}ne},
  \citenamefont {Ishibashi}, \citenamefont {Buhmann},\ and\ \citenamefont
  {Molenkamp}}]{bocquillon2017gapless}%
  \BibitemOpen
  \bibfield  {author} {\bibinfo {author} {\bibfnamefont {E.}~\bibnamefont
  {Bocquillon}}, \bibinfo {author} {\bibfnamefont {R.~S.}\ \bibnamefont
  {Deacon}}, \bibinfo {author} {\bibfnamefont {J.}~\bibnamefont {Wiedenmann}},
  \bibinfo {author} {\bibfnamefont {P.}~\bibnamefont {Leubner}}, \bibinfo
  {author} {\bibfnamefont {T.~M.}\ \bibnamefont {Klapwijk}}, \bibinfo {author}
  {\bibfnamefont {C.}~\bibnamefont {Br{\"u}ne}}, \bibinfo {author}
  {\bibfnamefont {K.}~\bibnamefont {Ishibashi}}, \bibinfo {author}
  {\bibfnamefont {H.}~\bibnamefont {Buhmann}},\ and\ \bibinfo {author}
  {\bibfnamefont {L.~W.}\ \bibnamefont {Molenkamp}},\ }\bibfield  {title}
  {\bibinfo {title} {Gapless andreev bound states in the quantum spin hall
  insulator hgte},\ }\href {https://doi.org/10.1038/nnano.2016.159} {\bibfield
  {journal} {\bibinfo  {journal} {Nature Nanotechnology}\ }\textbf {\bibinfo
  {volume} {12}},\ \bibinfo {pages} {137} (\bibinfo {year} {2017})}\BibitemShut
  {NoStop}%
\bibitem [{\citenamefont {Li}\ \emph {et~al.}(2018{\natexlab{b}})\citenamefont
  {Li}, \citenamefont {de~Boer}, \citenamefont {de~Ronde}, \citenamefont
  {Ramankutty}, \citenamefont {van Heumen}, \citenamefont {Huang},
  \citenamefont {de~Visser}, \citenamefont {Golubov}, \citenamefont {Golden},\
  and\ \citenamefont {Brinkman}}]{li2018NM}%
  \BibitemOpen
  \bibfield  {author} {\bibinfo {author} {\bibfnamefont {C.}~\bibnamefont
  {Li}}, \bibinfo {author} {\bibfnamefont {J.~C.}\ \bibnamefont {de~Boer}},
  \bibinfo {author} {\bibfnamefont {B.}~\bibnamefont {de~Ronde}}, \bibinfo
  {author} {\bibfnamefont {S.~V.}\ \bibnamefont {Ramankutty}}, \bibinfo
  {author} {\bibfnamefont {E.}~\bibnamefont {van Heumen}}, \bibinfo {author}
  {\bibfnamefont {Y.}~\bibnamefont {Huang}}, \bibinfo {author} {\bibfnamefont
  {A.}~\bibnamefont {de~Visser}}, \bibinfo {author} {\bibfnamefont {A.~A.}\
  \bibnamefont {Golubov}}, \bibinfo {author} {\bibfnamefont {M.~S.}\
  \bibnamefont {Golden}},\ and\ \bibinfo {author} {\bibfnamefont
  {A.}~\bibnamefont {Brinkman}},\ }\bibfield  {title} {\bibinfo {title}
  {4$\pi$-periodic andreev bound states in a dirac semimetal},\ }\href
  {https://doi.org/10.1038/s41563-018-0158-6} {\bibfield  {journal} {\bibinfo
  {journal} {Nature Materials}\ }\textbf {\bibinfo {volume} {17}},\ \bibinfo
  {pages} {875} (\bibinfo {year} {2018}{\natexlab{b}})}\BibitemShut {NoStop}%
\bibitem [{\citenamefont {Wang}\ \emph {et~al.}(2018)\citenamefont {Wang},
  \citenamefont {Li}, \citenamefont {Li}, \citenamefont {Liao}, \citenamefont
  {Brinkman},\ and\ \citenamefont {Yu}}]{wang2018Shapiro}%
  \BibitemOpen
  \bibfield  {author} {\bibinfo {author} {\bibfnamefont {A.-Q.}\ \bibnamefont
  {Wang}}, \bibinfo {author} {\bibfnamefont {C.-Z.}\ \bibnamefont {Li}},
  \bibinfo {author} {\bibfnamefont {C.}~\bibnamefont {Li}}, \bibinfo {author}
  {\bibfnamefont {Z.-M.}\ \bibnamefont {Liao}}, \bibinfo {author}
  {\bibfnamefont {A.}~\bibnamefont {Brinkman}},\ and\ \bibinfo {author}
  {\bibfnamefont {D.-P.}\ \bibnamefont {Yu}},\ }\bibfield  {title} {\bibinfo
  {title} {$4\ensuremath{\pi}$-periodic supercurrent from surface states in
  ${\mathrm{cd}}_{3}{\mathrm{as}}_{2}$ nanowire-based josephson junctions},\
  }\href {https://doi.org/10.1103/PhysRevLett.121.237701} {\bibfield  {journal}
  {\bibinfo  {journal} {Phys. Rev. Lett.}\ }\textbf {\bibinfo {volume} {121}},\
  \bibinfo {pages} {237701} (\bibinfo {year} {2018})}\BibitemShut {NoStop}%
\bibitem [{\citenamefont {Sch{\"u}ffelgen}\ \emph {et~al.}(2019)\citenamefont
  {Sch{\"u}ffelgen}, \citenamefont {Rosenbach}, \citenamefont {Li},
  \citenamefont {Schmitt}, \citenamefont {Schleenvoigt}, \citenamefont {Jalil},
  \citenamefont {Schmitt}, \citenamefont {K{\"o}lzer}, \citenamefont {Wang},
  \citenamefont {Bennemann}, \citenamefont {Parlak}, \citenamefont {Kibkalo},
  \citenamefont {Trellenkamp}, \citenamefont {Grap}, \citenamefont {Meertens},
  \citenamefont {Luysberg}, \citenamefont {Mussler}, \citenamefont
  {Berenschot}, \citenamefont {Tas}, \citenamefont {Golubov}, \citenamefont
  {Brinkman}, \citenamefont {Sch{\"a}pers},\ and\ \citenamefont
  {Gr{\"u}tzmacher}}]{Schuffelgen2019}%
  \BibitemOpen
  \bibfield  {author} {\bibinfo {author} {\bibfnamefont {P.}~\bibnamefont
  {Sch{\"u}ffelgen}}, \bibinfo {author} {\bibfnamefont {D.}~\bibnamefont
  {Rosenbach}}, \bibinfo {author} {\bibfnamefont {C.}~\bibnamefont {Li}},
  \bibinfo {author} {\bibfnamefont {T.~W.}\ \bibnamefont {Schmitt}}, \bibinfo
  {author} {\bibfnamefont {M.}~\bibnamefont {Schleenvoigt}}, \bibinfo {author}
  {\bibfnamefont {A.~R.}\ \bibnamefont {Jalil}}, \bibinfo {author}
  {\bibfnamefont {S.}~\bibnamefont {Schmitt}}, \bibinfo {author} {\bibfnamefont
  {J.}~\bibnamefont {K{\"o}lzer}}, \bibinfo {author} {\bibfnamefont
  {M.}~\bibnamefont {Wang}}, \bibinfo {author} {\bibfnamefont {B.}~\bibnamefont
  {Bennemann}}, \bibinfo {author} {\bibfnamefont {U.}~\bibnamefont {Parlak}},
  \bibinfo {author} {\bibfnamefont {L.}~\bibnamefont {Kibkalo}}, \bibinfo
  {author} {\bibfnamefont {S.}~\bibnamefont {Trellenkamp}}, \bibinfo {author}
  {\bibfnamefont {T.}~\bibnamefont {Grap}}, \bibinfo {author} {\bibfnamefont
  {D.}~\bibnamefont {Meertens}}, \bibinfo {author} {\bibfnamefont
  {M.}~\bibnamefont {Luysberg}}, \bibinfo {author} {\bibfnamefont
  {G.}~\bibnamefont {Mussler}}, \bibinfo {author} {\bibfnamefont
  {E.}~\bibnamefont {Berenschot}}, \bibinfo {author} {\bibfnamefont
  {N.}~\bibnamefont {Tas}}, \bibinfo {author} {\bibfnamefont {A.~A.}\
  \bibnamefont {Golubov}}, \bibinfo {author} {\bibfnamefont {A.}~\bibnamefont
  {Brinkman}}, \bibinfo {author} {\bibfnamefont {T.}~\bibnamefont
  {Sch{\"a}pers}},\ and\ \bibinfo {author} {\bibfnamefont {D.}~\bibnamefont
  {Gr{\"u}tzmacher}},\ }\bibfield  {title} {\bibinfo {title} {Selective area
  growth and stencil lithography for in situ fabricated quantum devices},\
  }\href {https://doi.org/10.1038/s41565-019-0506-y} {\bibfield  {journal}
  {\bibinfo  {journal} {Nature Nanotechnology}\ }\textbf {\bibinfo {volume}
  {14}},\ \bibinfo {pages} {825} (\bibinfo {year} {2019})}\BibitemShut
  {NoStop}%
\bibitem [{\citenamefont {Le~Calvez}\ \emph {et~al.}(2019)\citenamefont
  {Le~Calvez}, \citenamefont {Veyrat}, \citenamefont {Gay}, \citenamefont
  {Plaindoux}, \citenamefont {Winkelmann}, \citenamefont {Courtois},\ and\
  \citenamefont {Sac{\'e}p{\'e}}}]{LeCalvez2019}%
  \BibitemOpen
  \bibfield  {author} {\bibinfo {author} {\bibfnamefont {K.}~\bibnamefont
  {Le~Calvez}}, \bibinfo {author} {\bibfnamefont {L.}~\bibnamefont {Veyrat}},
  \bibinfo {author} {\bibfnamefont {F.}~\bibnamefont {Gay}}, \bibinfo {author}
  {\bibfnamefont {P.}~\bibnamefont {Plaindoux}}, \bibinfo {author}
  {\bibfnamefont {C.~B.}\ \bibnamefont {Winkelmann}}, \bibinfo {author}
  {\bibfnamefont {H.}~\bibnamefont {Courtois}},\ and\ \bibinfo {author}
  {\bibfnamefont {B.}~\bibnamefont {Sac{\'e}p{\'e}}},\ }\bibfield  {title}
  {\bibinfo {title} {Joule overheating poisons the fractional ac josephson
  effect in topological josephson junctions},\ }\href
  {https://doi.org/10.1038/s42005-018-0100-x} {\bibfield  {journal} {\bibinfo
  {journal} {Communications Physics}\ }\textbf {\bibinfo {volume} {2}},\
  \bibinfo {pages} {4} (\bibinfo {year} {2019})}\BibitemShut {NoStop}%
\bibitem [{\citenamefont {Rosenbach}\ \emph {et~al.}(2021)\citenamefont
  {Rosenbach}, \citenamefont {Schmitt}, \citenamefont {Schüffelgen},
  \citenamefont {Stehno}, \citenamefont {Li}, \citenamefont {Schleenvoigt},
  \citenamefont {Jalil}, \citenamefont {Mussler}, \citenamefont {Neumann},
  \citenamefont {Trellenkamp}, \citenamefont {Golubov}, \citenamefont
  {Brinkman}, \citenamefont {Grützmacher},\ and\ \citenamefont
  {Schäpers}}]{rosenbach2021}%
  \BibitemOpen
  \bibfield  {author} {\bibinfo {author} {\bibfnamefont {D.}~\bibnamefont
  {Rosenbach}}, \bibinfo {author} {\bibfnamefont {T.~W.}\ \bibnamefont
  {Schmitt}}, \bibinfo {author} {\bibfnamefont {P.}~\bibnamefont
  {Schüffelgen}}, \bibinfo {author} {\bibfnamefont {M.~P.}\ \bibnamefont
  {Stehno}}, \bibinfo {author} {\bibfnamefont {C.}~\bibnamefont {Li}}, \bibinfo
  {author} {\bibfnamefont {M.}~\bibnamefont {Schleenvoigt}}, \bibinfo {author}
  {\bibfnamefont {A.~R.}\ \bibnamefont {Jalil}}, \bibinfo {author}
  {\bibfnamefont {G.}~\bibnamefont {Mussler}}, \bibinfo {author} {\bibfnamefont
  {E.}~\bibnamefont {Neumann}}, \bibinfo {author} {\bibfnamefont
  {S.}~\bibnamefont {Trellenkamp}}, \bibinfo {author} {\bibfnamefont {A.~A.}\
  \bibnamefont {Golubov}}, \bibinfo {author} {\bibfnamefont {A.}~\bibnamefont
  {Brinkman}}, \bibinfo {author} {\bibfnamefont {D.}~\bibnamefont
  {Grützmacher}},\ and\ \bibinfo {author} {\bibfnamefont {T.}~\bibnamefont
  {Schäpers}},\ }\bibfield  {title} {\bibinfo {title} {Reappearance of first
  shapiro step in narrow topological josephson junctions},\ }\href
  {https://doi.org/10.1126/sciadv.abf1854} {\bibfield  {journal} {\bibinfo
  {journal} {Science Advances}\ }\textbf {\bibinfo {volume} {7}},\ \bibinfo
  {pages} {eabf1854} (\bibinfo {year} {2021})}\BibitemShut {NoStop}%
\bibitem [{sup()}]{supp}%
  \BibitemOpen
  \href@noop {} {}\bibinfo {howpublished} {See the Supplemental Materials for
  the detailed derivation of the semiclassical equations for the Josephson
  phase and the pseudo spin.}\BibitemShut {Stop}%
\bibitem [{\citenamefont {Kitaev}(2003)}]{kitaev2003anyon}%
  \BibitemOpen
  \bibfield  {author} {\bibinfo {author} {\bibfnamefont {A.~Y.}\ \bibnamefont
  {Kitaev}},\ }\bibfield  {title} {\bibinfo {title} {Fault-tolerant quantum
  computation by anyons},\ }\href
  {https://doi.org/10.1016/S0003-4916(02)00018-0} {\bibfield  {journal}
  {\bibinfo  {journal} {Annals of Physics}\ }\textbf {\bibinfo {volume}
  {303}},\ \bibinfo {pages} {2} (\bibinfo {year} {2003})}\BibitemShut {NoStop}%
\bibitem [{\citenamefont {Lutchyn}\ \emph {et~al.}(2010)\citenamefont
  {Lutchyn}, \citenamefont {Sau},\ and\ \citenamefont
  {Das~Sarma}}]{lutchyn2010TSC1d}%
  \BibitemOpen
  \bibfield  {author} {\bibinfo {author} {\bibfnamefont {R.~M.}\ \bibnamefont
  {Lutchyn}}, \bibinfo {author} {\bibfnamefont {J.~D.}\ \bibnamefont {Sau}},\
  and\ \bibinfo {author} {\bibfnamefont {S.}~\bibnamefont {Das~Sarma}},\
  }\bibfield  {title} {\bibinfo {title} {Majorana fermions and a topological
  phase transition in semiconductor-superconductor heterostructures},\ }\href
  {https://doi.org/10.1103/PhysRevLett.105.077001} {\bibfield  {journal}
  {\bibinfo  {journal} {Phys. Rev. Lett.}\ }\textbf {\bibinfo {volume} {105}},\
  \bibinfo {pages} {077001} (\bibinfo {year} {2010})}\BibitemShut {NoStop}%
\bibitem [{\citenamefont {Oreg}\ \emph {et~al.}(2010)\citenamefont {Oreg},
  \citenamefont {Refael},\ and\ \citenamefont {von Oppen}}]{oreg2010TSC1d}%
  \BibitemOpen
  \bibfield  {author} {\bibinfo {author} {\bibfnamefont {Y.}~\bibnamefont
  {Oreg}}, \bibinfo {author} {\bibfnamefont {G.}~\bibnamefont {Refael}},\ and\
  \bibinfo {author} {\bibfnamefont {F.}~\bibnamefont {von Oppen}},\ }\bibfield
  {title} {\bibinfo {title} {Helical liquids and majorana bound states in
  quantum wires},\ }\href {https://doi.org/10.1103/PhysRevLett.105.177002}
  {\bibfield  {journal} {\bibinfo  {journal} {Phys. Rev. Lett.}\ }\textbf
  {\bibinfo {volume} {105}},\ \bibinfo {pages} {177002} (\bibinfo {year}
  {2010})}\BibitemShut {NoStop}%
\bibitem [{\citenamefont {Josephson}(1962)}]{josephson1962}%
  \BibitemOpen
  \bibfield  {author} {\bibinfo {author} {\bibfnamefont {B.~D.}\ \bibnamefont
  {Josephson}},\ }\bibfield  {title} {\bibinfo {title} {Possible new effects in
  superconductive tunnelling},\ }\href@noop {} {\bibfield  {journal} {\bibinfo
  {journal} {Physics Letters}\ }\textbf {\bibinfo {volume} {1}},\ \bibinfo
  {pages} {251} (\bibinfo {year} {1962})}\BibitemShut {NoStop}%
\bibitem [{\citenamefont {Wen}(2004)}]{wen2004}%
  \BibitemOpen
  \bibfield  {author} {\bibinfo {author} {\bibfnamefont {X.-G.}\ \bibnamefont
  {Wen}},\ }\bibfield  {title} {\bibinfo {title} {\emph{Quantum Field Theory of
  Many-body Systems from the Origin of Sound to an Origin of Light and
  Electrons}},\ }\href@noop {} {\bibfield  {journal} {\bibinfo  {journal}
  {\emph{ Quantum Field Theory of Many-body Systems. } Oxford University Press
  Inc., New York}\ } (\bibinfo {year} {2004})}\BibitemShut {NoStop}%
\bibitem [{\citenamefont {Altland}\ and\ \citenamefont
  {Simons}(2010)}]{Altland2010condensed}%
  \BibitemOpen
  \bibfield  {author} {\bibinfo {author} {\bibfnamefont {A.}~\bibnamefont
  {Altland}}\ and\ \bibinfo {author} {\bibfnamefont {B.~D.}\ \bibnamefont
  {Simons}},\ }\href@noop {} {\emph {\bibinfo {title} {Condensed matter field
  theory}}}\ (\bibinfo  {publisher} {Cambridge university press},\ \bibinfo
  {year} {2010})\BibitemShut {NoStop}%
\bibitem [{\citenamefont {Caldeira}\ and\ \citenamefont
  {Leggett}(1981)}]{Caldeira1981TLS}%
  \BibitemOpen
  \bibfield  {author} {\bibinfo {author} {\bibfnamefont {A.~O.}\ \bibnamefont
  {Caldeira}}\ and\ \bibinfo {author} {\bibfnamefont {A.~J.}\ \bibnamefont
  {Leggett}},\ }\bibfield  {title} {\bibinfo {title} {Influence of dissipation
  on quantum tunneling in macroscopic systems},\ }\href
  {https://doi.org/10.1103/PhysRevLett.46.211} {\bibfield  {journal} {\bibinfo
  {journal} {Phys. Rev. Lett.}\ }\textbf {\bibinfo {volume} {46}},\ \bibinfo
  {pages} {211} (\bibinfo {year} {1981})}\BibitemShut {NoStop}%
\bibitem [{\citenamefont {Caldeira}\ and\ \citenamefont
  {Leggett}(1983)}]{Caldeira1983Brownian}%
  \BibitemOpen
  \bibfield  {author} {\bibinfo {author} {\bibfnamefont {A.}~\bibnamefont
  {Caldeira}}\ and\ \bibinfo {author} {\bibfnamefont {A.}~\bibnamefont
  {Leggett}},\ }\bibfield  {title} {\bibinfo {title} {Path integral approach to
  quantum brownian motion},\ }\href
  {https://doi.org/https://doi.org/10.1016/0378-4371(83)90013-4} {\bibfield
  {journal} {\bibinfo  {journal} {Physica A: Statistical Mechanics and its
  Applications}\ }\textbf {\bibinfo {volume} {121}},\ \bibinfo {pages} {587}
  (\bibinfo {year} {1983})}\BibitemShut {NoStop}%
\bibitem [{\citenamefont {Leggett}\ \emph {et~al.}(1987)\citenamefont
  {Leggett}, \citenamefont {Chakravarty}, \citenamefont {Dorsey}, \citenamefont
  {Fisher}, \citenamefont {Garg},\ and\ \citenamefont
  {Zwerger}}]{Leggett1987RMP}%
  \BibitemOpen
  \bibfield  {author} {\bibinfo {author} {\bibfnamefont {A.~J.}\ \bibnamefont
  {Leggett}}, \bibinfo {author} {\bibfnamefont {S.}~\bibnamefont
  {Chakravarty}}, \bibinfo {author} {\bibfnamefont {A.~T.}\ \bibnamefont
  {Dorsey}}, \bibinfo {author} {\bibfnamefont {M.~P.~A.}\ \bibnamefont
  {Fisher}}, \bibinfo {author} {\bibfnamefont {A.}~\bibnamefont {Garg}},\ and\
  \bibinfo {author} {\bibfnamefont {W.}~\bibnamefont {Zwerger}},\ }\bibfield
  {title} {\bibinfo {title} {Dynamics of the dissipative two-state system},\
  }\href {https://doi.org/10.1103/RevModPhys.59.1} {\bibfield  {journal}
  {\bibinfo  {journal} {Rev. Mod. Phys.}\ }\textbf {\bibinfo {volume} {59}},\
  \bibinfo {pages} {1} (\bibinfo {year} {1987})}\BibitemShut {NoStop}%
\bibitem [{\citenamefont {Alicea}\ \emph {et~al.}(2011)\citenamefont {Alicea},
  \citenamefont {Oreg}, \citenamefont {Refael}, \citenamefont {von Oppen},\
  and\ \citenamefont {Fisher}}]{alicea2011non}%
  \BibitemOpen
  \bibfield  {author} {\bibinfo {author} {\bibfnamefont {J.}~\bibnamefont
  {Alicea}}, \bibinfo {author} {\bibfnamefont {Y.}~\bibnamefont {Oreg}},
  \bibinfo {author} {\bibfnamefont {G.}~\bibnamefont {Refael}}, \bibinfo
  {author} {\bibfnamefont {F.}~\bibnamefont {von Oppen}},\ and\ \bibinfo
  {author} {\bibfnamefont {M.~P.~A.}\ \bibnamefont {Fisher}},\ }\bibfield
  {title} {\bibinfo {title} {Non-abelian statistics and topological quantum
  information processing in 1d wire networks},\ }\href
  {https://doi.org/10.1038/nphys1915} {\bibfield  {journal} {\bibinfo
  {journal} {Nature Physics}\ }\textbf {\bibinfo {volume} {7}},\ \bibinfo
  {pages} {412} (\bibinfo {year} {2011})}\BibitemShut {NoStop}%
\bibitem [{\citenamefont {Shevchenko}\ \emph {et~al.}(2010)\citenamefont
  {Shevchenko}, \citenamefont {Ashhab},\ and\ \citenamefont
  {Nori}}]{shevchenko2010LZS}%
  \BibitemOpen
  \bibfield  {author} {\bibinfo {author} {\bibfnamefont {S.~N.}\ \bibnamefont
  {Shevchenko}}, \bibinfo {author} {\bibfnamefont {S.}~\bibnamefont {Ashhab}},\
  and\ \bibinfo {author} {\bibfnamefont {F.}~\bibnamefont {Nori}},\ }\bibfield
  {title} {\bibinfo {title} {Landau--zener--st{\"u}ckelberg interferometry},\
  }\href@noop {} {\bibfield  {journal} {\bibinfo  {journal} {Physics Reports}\
  }\textbf {\bibinfo {volume} {492}},\ \bibinfo {pages} {1} (\bibinfo {year}
  {2010})}\BibitemShut {NoStop}%
\bibitem [{\citenamefont {Feng}\ \emph {et~al.}(2020)\citenamefont {Feng},
  \citenamefont {Huang}, \citenamefont {Wang},\ and\ \citenamefont
  {Niu}}]{feng2020radiation}%
  \BibitemOpen
  \bibfield  {author} {\bibinfo {author} {\bibfnamefont {J.-J.}\ \bibnamefont
  {Feng}}, \bibinfo {author} {\bibfnamefont {Z.}~\bibnamefont {Huang}},
  \bibinfo {author} {\bibfnamefont {Z.}~\bibnamefont {Wang}},\ and\ \bibinfo
  {author} {\bibfnamefont {Q.}~\bibnamefont {Niu}},\ }\bibfield  {title}
  {\bibinfo {title} {Josephson radiation from nonlinear dynamics of majorana
  zero modes},\ }\href {https://doi.org/10.1103/PhysRevB.101.180504} {\bibfield
   {journal} {\bibinfo  {journal} {Phys. Rev. B}\ }\textbf {\bibinfo {volume}
  {101}},\ \bibinfo {pages} {180504} (\bibinfo {year} {2020})}\BibitemShut
  {NoStop}%
\bibitem [{\citenamefont {Gilbert}(2004)}]{Gilbert2004}%
  \BibitemOpen
  \bibfield  {author} {\bibinfo {author} {\bibfnamefont {T.}~\bibnamefont
  {Gilbert}},\ }\bibfield  {title} {\bibinfo {title} {A phenomenological theory
  of damping in ferromagnetic materials},\ }\href
  {https://doi.org/10.1109/TMAG.2004.836740} {\bibfield  {journal} {\bibinfo
  {journal} {IEEE Transactions on Magnetics}\ }\textbf {\bibinfo {volume}
  {40}},\ \bibinfo {pages} {3443} (\bibinfo {year} {2004})}\BibitemShut
  {NoStop}%
\bibitem [{\citenamefont {Sanders}\ \emph {et~al.}(2007)\citenamefont
  {Sanders}, \citenamefont {Verhulst},\ and\ \citenamefont
  {Murdock}}]{sanders2007averaging}%
  \BibitemOpen
  \bibfield  {author} {\bibinfo {author} {\bibfnamefont {J.~A.}\ \bibnamefont
  {Sanders}}, \bibinfo {author} {\bibfnamefont {F.}~\bibnamefont {Verhulst}},\
  and\ \bibinfo {author} {\bibfnamefont {J.}~\bibnamefont {Murdock}},\
  }\href@noop {} {\emph {\bibinfo {title} {Averaging methods in nonlinear
  dynamical systems}}},\ Vol.~\bibinfo {volume} {59}\ (\bibinfo  {publisher}
  {Springer},\ \bibinfo {year} {2007})\BibitemShut {NoStop}%
\end{thebibliography}%

\end{document}